\documentclass[12pt]{iopart}

\usepackage{amscd,amssymb}

\usepackage{epsfig}
\usepackage{amscd,amssymb}
\usepackage{graphicx,xspace}
\usepackage{pstricks}
\usepackage[left=3cm,top=3cm,right=3cm,bottom=3cm]{geometry}
\usepackage{color}
\definecolor{labelkey}{cmyk}{.4,.2,0,0}

\begin{document}

\newcommand \be  {\begin{equation}}
\newcommand \bea {\begin{eqnarray} \nonumber }
\newcommand \ee  {\end{equation}}
\newcommand \eea {\end{eqnarray}}

\title{Topology trivialization and large deviations for the minimum in the simplest random optimization.}

\vskip 0.2cm

\author{Yan V Fyodorov}

\address{School of Mathematical Sciences, Queen Mary University of London\\  London E1 4NS, United Kingdom}

\author{Pierre Le Doussal}

\address{CNRS-Laboratoire de Physique Th\'eorique de l'Ecole Normale Sup\'erieure\\
24 rue Lhomond, 75231 Paris
Cedex-France\footnote{LPTENS is a Unit\'e Propre du C.N.R.S.
associ\'ee \`a l'Ecole Normale Sup\'erieure et \`a l'Universit\'e Paris Sud}}

\begin{abstract}
Finding the global minimum of a cost function given by the sum of a quadratic and a linear form in $N$  real variables over $(N-1)-$ dimensional sphere is one of the simplest, yet paradigmatic problems in Optimization Theory known as the "trust region subproblem" or "constraint least square problem".
When both terms in the cost function are random this amounts to studying the ground state energy of the simplest spherical spin glass in a random magnetic field.  We first identify and study two distinct large-$N$ scaling regimes in which the linear term (magnetic field) leads to a gradual topology trivialization, i.e. reduction in the total number ${\cal N}_{tot}$ of critical (stationary) points in the cost function landscape. In the first regime ${\cal N}_{tot}$ remains of the order $N$
and the cost function (energy) has generically two almost degenerate minima with the Tracy-Widom (TW) statistics.
In the second regime the number of critical points is of the order of unity
 with a finite probability for a single minimum.
In that case the mean total number of extrema (minima and maxima)
of the cost function is given by the Laplace transform of the TW density, and the distribution of the global minimum energy is expected to take a universal scaling form
generalizing the TW law. Though the full form of that distribution is not yet known to us,
 one of its far tails can be inferred from the large deviation theory for the global minimum.
 In the rest of the paper we show how to use the replica method to obtain
 the probability density of the minimum energy in the large-deviation approximation
 by finding both the rate function and the leading pre-exponential factor.
\end{abstract}

\maketitle

\section{Introduction}
The problem of minimizing the quadratic over the sphere
\be\label{1}
 E_{min}(h)=\min_{|{\bf x}|=R} \left\{E_{h}({\bf x})\right\},\,\, E_{h}({\bf x})=-\frac{1}{2}{\bf x}^{T} H {\bf x}- {\bf h}^T{\bf x}, \quad  {\bf h},{\bf x}\in \mathbb{R}_N,
\ee
  plays important role in the Optimization Theory as it naturally arises at every step of iteration in a popular class of nonlinear optimization algorithms called "trust region methods" \cite{trust}. In a different incarnation it is known as the simplest representative of "constraint least square problems"\cite{QEP,Golub89}. For these reasons a lot of effort was devoted to developing effective numerical algorithms for its solution, especially for large dimensions, see \cite{Hager,Sorensen} and references therein. For ${\bf h}=0$ the problem is equivalent to finding the maximal eigenvalue of the $N\times N$ real symmetric matrix $H$ and in this sense straightforward both conceptually and numerically. The case ${\bf h}\ne 0$ is equivalent to a certain "quadratic eigenvalue problem" \cite{QEP} whose solution can be written in terms of the roots of the equation involving the resolvent of $H$, see \cite{FG} and equations (\ref{2}) and (\ref{3}) below, which makes investigating the properties of the minimum considerably more challenging. From the point of view of Statistical Mechanics the cost function  $E_{h}({\bf x})$ has a natural interpretation of the energy associated with a configuration ${\bf x}^T=(x_1,\ldots, x_N)$ of $N$ spin variables $x_i$, with $H$ standing for the spin interaction matrix and ${\bf h}$ for the magnetic field. In that context the constraint $|{\bf x}|=\sqrt{N}$ defines the so-called spherical spin model. Further assuming $H$ to be a random $N\times N$  matrix from the Gaussian Orthogonal Ensemble (GOE) defines the simplest spherical spin-glass model introduced and studied for $N\gg 1$ long ago by Kosterlitz, Thouless and Jones \cite{KTJ}, and by many authors ever since, see e.g. chap. 4 of the book \cite{DG}. The statistics of the global energy minimum (the ground state) of such a spin glass for ${\bf h}=0$ is trivially related to the properties of the maximal eigenvalues of GOE matrices. The latter is by now well-studied in the random matrix theory (RMT) and given by the famous Tracy-Widom law \cite{TW} in the small-deviation regime, and by well known large-deviation functionals beyond that regime \cite{BDG,DM,MV,BorotNadal11,For1}. We also note that there exists close and fruitful relation between RMT large deviations functionals, spherical spin glasses, and the problem of counting minima and saddle points of large-dimensional disordered surfaces, see \cite{Fyo04,BD07,FyoWi07,FyNa12,Auf1,Auf2} and references therein, and the section 3 of the present paper.

Although thermodynamics of the model is simple and does not show such prominent features as replica-symmetry breaking,
dynamics for ${\bf h}=0$ is rich and has features of aging \cite{CD,ZKH,BDG}. That richness is attributed
to a relatively rich energy landscape topology due to presence of $2N$ stationary points in the landscape.
It was further noticed by Cugliandolo and Dean in \cite{CD1} that taking an arbitrary small $N-$independent
magnetic field ${\bf h}\ne 0$ trivializes the topology in the $N\to \infty$ limit by allowing only two stationary points to survive, the maximum and the minimum
(see sections 2 and 3 below for a detailed discussion). Such an abrupt restructuring of the landscape indeed was shown to result in washing out the aging effects 
for any finite value of the magnetic field \cite{CD1}. The first main goal of our paper is to provide a detailed, quantitative picture of the topology trivialization for large but finite $N\gg 1$. Namely, we will identify and study in some detail the existence of two nontrivial scaling regimes: $|{\bf h}|\sim N^{-1/2}$  and $|{\bf h}|\sim N^{-1/6}$. In the former the topology is still complex in the sense of existence of the order of $N$ stationary points. In the latter the number of stationary points is finite, gradually decreases with growing field and tends to just two, a minimum and a maximum, when $|{\bf h}| N^{1/6}\gg 1$. Note that extending the analysis of the present paper it can be demonstrated that essentially the same scenario of the topology trivialization takes place in a general spherical spinglass model with $p-$spin interaction in the scaling vicinity of the replica symmetry breaking point\cite{Fyolec}.

 Having understood in some detail the picture of gradual landscape topology trivialization we then address the question of statistics of the global energy minimum in the presence of a nonzero random magnetic field. The question is not trivial and, to the best of our knowledge, has not been much studied. The difficulty is that for ${\bf h}\ne 0$ the relation to properties of random matrices is less direct, see the next section for a discussion, and the powerful RMT tools do not seem to be of obvious utility.

    To that end, a simple perturbation theory insights suggest that in the first scaling regime $|{\bf h}|\sim N^{-1/2}$ the magnetic field is too small to modify the Tracy-Widom statistics of the global minimum. In contrast, fields of the order $|{\bf h}|\sim N^{-1/6}$ do modify statistics of the extrema, and we expect the distribution of the minimum in that scaling regime to be given by a family of universal laws generalizing the TW distribution and containing the latter as a limiting case. Though finding the explicit description of the family remains an outstanding challenge one can get some insights from the side of {\it large deviations} pertinent to the case of fields $|h|\sim O(1)$.  From that angle the second main goal of this paper is to show that the explicit form of the probability density for the minimum can still be found in the large-deviation regime in some range around its typical value. This can be done in the framework of the replica trick which we will use in two alternative ways. Following the first way  one extracts the Legendre transform of the large deviation rate function from analysing the $n-$dependence of the moments of the partition function.  This is very close to the method of Parisi $\&$ Rizzo\cite{PR08,PR09,PR10} employed in their recent studies of  large deviations of free energy of the Sherrington-Kirkpatrick model, though we concentrate for our case on the zero-temperature limit and aim to derive the full large-deviation rate function rather than its perturbative expansion. That expression in the limit  of vanishing field successfully reproduces the known RMT large-deviation results.
    We also show that the method is capable of producing the leading pre-exponential factor by taking into account the Gaussian fluctuations around the saddle-point solution with the help of de-Almeida-Thouless\cite{AT}-inspired fluctuation determinant analysis.  In the limit ${\bf h} \to 0$ this factor is found to reproduce correctly the structure of the known RMT pre-factors,  up to a global factor of $2$ (accordingly, the naive zero field limit of our expression is exactly twice the asymptotics of the Tracy-Widom distribution in the small deviations regime).
    We will discuss a possible scenario behind such a mismatch. Thus though our large-deviation calculations provide a hint that the TW distribution may be tackled using replica, recovering the full expression  remains a considerable challenge. The calculation also allows us to predict the form of one of the tails of the (presumably universal) distribution for the energy minimum at magnetic fields $|h|\sim N^{-1/6}$.

  Finally, in the last section of the paper we suggest an alternative method allowing us to arrive to the same large-deviation rate function by directly addressing the probability density for the ground state in the replica limit $n\to 0$. It seems to be new to the best of our knowledge. We hope that the method, after due modification, may prove to be useful for studying more complicated optimization problems, such as large deviation functionals of the ground states in systems which show broken replica symmetry like more general spherical spin glasses \cite{CS} or related disordered models \cite{FS,FB,KPZreplica1,KPZreplica2,KPZreplica3,KPZreplica4}.

The paper has the following structure. We begin with outlining the exact formal solution for the minimization problem (\ref{1}) in terms of the resolvent of the matrix $H$ and briefly discuss how the position of the typical minimum can be inferred from a simple RMT consideration. We also use perturbation theory to relate the statistics of the minimum for very small magnetic fields to some interesting objects in the random matrix theory and further identify two nontrivial scaling regimes for the magnetic field related to the gradual topology trivialization. In essence, those two regimes stem from the existence of the RMT "bulk" and "edge" spectral regimes. Then we provide the explicit calculation of the mean number of critical points in the first scaling regime, show that under such a scaling that number is of the order of $N$ and becomes of the order of unity when approaching the second scaling regime. The same calculation is extended to the second scaling regime, where also relate the mean number of minima to the Tracy-Widom density. In the rest of the paper we describe two versions of the replica trick used to derive the large-deviation rate for the minimal energy in two alternative ways, and also show how to take into account the fluctuation determinant contribution to find the leading pre-exponential factor. Finally, in the conclusion section we formulate a few open problems stemming from our research.

\section{Lagrange multiplier minimization. Relation to RMT in perturbative and small deviation regimes.}
We begin with outlining the exact formal solution for the minimization problem (\ref{1}) given originally in \cite{FG}. Applying the Lagrange multiplier method to (\ref{1}) by adding to the cost function the term $t({\bf x}^T{\bf x}-R^2)$ and minimizing yields in the standard way the argmin ${\bf x}_*$ of the cost function as ${\bf x}_*=(t_*-H)^{-1}{\bf h}$ where the multiplier $t_*$ is the {\it maximal} solution of the following secular equation:
\be\label{2}
R^2={\bf h}^T\frac{1}{(t-H)^2}{\bf h}=\sum_{j=1}^N \frac{w_j}{(t-\lambda_j)^2}, \quad  w_j=({\bf h}^T{\bf e}_j)\left({\bf e}_j^T{\bf h}\right)
\ee
where ${\bf e}_j$  are the orthonormal eigenvectors of $H$ and $\lambda_j$ denote the corresponding real eigenvalues.

For a generic situation the vector ${\bf h}$ is not parallel to one of the eigenvectors and one can show  that $t_*>\mbox{max}_j\{\lambda_j\}$ \cite{FG}. The minimal value of the cost function is then given by
\be\label{3}
E_{min}(h)=-\frac{1}{2}\left(R^2 t_*+{\bf h}^T\frac{1}{(t_*-H)}{\bf h}\right)=-\frac{1}{2}\left(R^2 t_*+\sum_{j=1}^N \frac{w_j}{t_*-\lambda_j}\right)
\ee

 In the remainder of this paper we consider $N\times N$ matrices $H\in GOE$ distributed according to
the weight $ {\cal P}(H)\propto \exp{-\frac{N}{4J^2}\mbox{\small Tr}H^2}$, i.e entries $H_{ij}=H_{ji}$ are independent mean zero Gaussian real variables with variances $<H_{ij}^2>=J^2/N$ for $i <j$,
and $<H_{ii}^2>=2 J^2/N$. We treat also the components $h_i$ of the field ${\bf h}$ as independent, identically distributed random Gaussian variables with zero mean and the variance  $\langle h^2_i\rangle=\sigma^2$, and use the spherical model constraint $R^2=N$. To this end we would like to note that had we replaced the {\it random field} term ${\bf h}^T{\bf x}$ in (\ref{1}) with the {\it random anisotropy} term $({\bf h}^T{\bf x})^2$ the resulting energy function could be written $E_{h}({\bf x})=-\frac{1}{2}{\bf x}^{T} \left(H+2{\bf h}\otimes {\bf h}^T \right){\bf x}$. The minimization problem would then amount to studying the maximal eigenvalue of a rank-one random perturbation of GOE, which
attracted a considerable interest recently, see e.g. \cite{Pe,FPe,BGM1,BFF} and whose large deviation functional is known explicitly\cite{BGM2}.
In contrast, the problem with magnetic field is not a simple eigenvalue problem but is equivalent to a much less studied class of quadratic eigenvalue problems \cite{QEP}. In particular, it is straightforward  to show that the secular equation (\ref{2}) for the Lagrange multipliers $t$ can be rewritten as  $\det{\left[R^{2}\,(t-H)^2-{\bf h}\otimes {\bf h}^T \right]}=0$. Analysing some features of our problem from that perspective could be an interesting line of research in its own sake but is not pursued in the present paper.

 What is simple to understand is why generically for $|{\bf h}|\sim \sigma$ of order of unity and large $N\gg 1$ the secular equation (\ref{2}) should have only two solutions, as well as to find the typical values of $t$ and the minimum energy in that case \cite{CD1}. First we recall that the typical spectrum of GOE matrices in the chosen normalisation is located in the interval $(-2J,2J)$. This implies that a typical separation $\Delta$ between neighbouring eigenvalues in that interval is of the order of $\Delta\sim JN^{-1}$. We immediately see that for any $t\in (-2J,2J)$ the right-hand side
in (\ref{2}) is typically of the order of $\sigma^2 \Delta^{-2}\sim (\sigma/J)^2  N^2$, whereas the left hand side is $R^2=N$. Therefore only for small magnetic fields of the order $\sigma\sim J N^{-1/2}$ such equation may have its solution $t$ in the interval $t\in (-2J,2J)$.
 In the next section we will find the mean number ${\cal N}_{tot}$ of solution in such a regime as a function of parameter $\gamma\sim N\sigma^2/J^2=O(1)$.
 We will find that ${\cal N}_{tot}$ is proportional to $N$ and gradually decreases with growth of $\gamma$ reflecting the phenomenon of topology trivialization. When the magnetic field reaches the scale $\sigma/J\sim N^{-1/6}$ the mean number of solutions in the interval $t\in (-2J,2J)$ is of the order of unity, and eventually, for $\sigma/J=O(1)$ there will be typically only two solutions, both outside that interval, with a single solution $t=t_*>2J$ corresponding to the energy {\it minimum}, and similarly another one with $t<-2J$ corresponding to the energy {\it maximum}.

To find the typical values of the Lagrange multiplier $t_*$ and of the minimum energy $E_{min}(h)$ for $\sigma=O(1)$ one may then take into account that that linear and quadratic terms in the cost function are not correlated and argue that the typical value of $t_*$ can be obtained by replacing (\ref{2}) with its ensemble averaged version:
 \be\label{4}
1=\sigma^2 \int_{-2J}^{2J} \frac{\rho_{sc}(\lambda)}{(t_*-\lambda)^2}\,d\lambda,\quad \rho_{sc}(\lambda)=\frac{1}{2\pi J^2}\sqrt{4J^2-\lambda^2}
\ee
Here we  used that the profile of the mean eigenvalue density in the interval $(-2J,2J)$ is given by the semicircular law $\rho_{sc}(\lambda)=\lim_{N\to \infty}\frac{1}{N}\left\langle\sum_j\delta(\lambda-\lambda_j)\right\rangle_{H}$, with $\lambda_j$ being the $N$ eigenvalues of $H$ and brackets standing for the ensemble averaging.  As a typical maximal Lagrange multiplier  $t_*>\lambda^{(typ)}_{max}=2J$, the integral in the right-hand side can be shown to be equal to $\frac{1}{2J^2}\left(\frac{t_*}{\sqrt{t_*^2-4J^2}}-1\right)$. Solving then the resulting equation and applying similar treatment to (\ref{3}) one finds after simple manipulations \cite{CD1}:
\be\label{5}
t^{(typ)}_*=\frac{\sigma^2+2J^2}{\sqrt{\sigma^2+J^2}}, \quad  E^{(typ)}_{min}(\sigma)=-N\sqrt{\sigma^2+J^2}
\ee
We will see later on in the paper that $E^{(typ)}_{min}(\sigma)$ is indeed both the typical and the average value of the ground state energy of the spherical spin glass as given by the replica trick.

Note that for  ${\bf h}=0$ to each solution $t=\lambda_i$ of the stationarity equation  (\ref{2}) corresponds exactly two different critical points of the cost function landscape with the same energy as changing ${\bf x}\to -{\bf x}$ does not change the cost function (\ref{1}). Thus we must have altogether $2N$ critical points. As for vanishing field we must have $t_*=\mbox{max}\{\lambda_1,\ldots,\lambda_N\}=\lambda_{max}$ it is reasonable to try to study the case of very weak fields by developing a perturbation theory around $\lambda_{max}$.

It can be done most conveniently by introducing a small parameter,  the typical scale $\sigma$ of the field, via formally defining $w_j=\sigma^2 \tilde{w}_j$, where now $\tilde{w}_j$ are considered to be of the order unity, and looking for the solution $t_*$ as a series in powers of $\sigma$. The straightforward manipulations yield for the first two nonvanishing terms of the expansion the following expression:
\be\label{per1}
t_*=\lambda_{max}+\sigma\sqrt{\frac{\tilde{w}_m}{N}}+\frac{\sigma^3}{2N}
\sqrt{\frac{\tilde{w}_m}{N}}\sum_{j\ne m}\frac{\tilde{w}_j}{(\lambda_{max}-\lambda_j)^2}+\ldots
\ee
where the sum goes over all $j=1,\ldots N$  excluding the terms with $\lambda_j=\lambda_{max}$, and we denoted
$\sigma^2\tilde{w}_m=({\bf h}^T{\bf e}_m)\left({\bf e}_m^T{\bf h}\right)$ where ${\bf e}_m$ stands for the eigenvector corresponding to the maximal eigenvalue. Further substituting (\ref{per1}) to (\ref{3}) yields a perturbative expansion for the minimal energy in the form
 \be\label{per2}
  E_{min}(h)=-\frac{1}{2}N\left\{\lambda_{max}+2\sigma\sqrt{\frac{\tilde{w}_m}{N}}+\frac{\sigma^2}{N}\sum_{j\ne m}\frac{\tilde{w}_j}{\lambda_{max}-\lambda_j}\right.
\ee
\[\left. -\frac{3\sigma^3}{2N}
\sqrt{\frac{\tilde{w}_m}{N}}\sum_{j\ne m}\frac{\tilde{w}_j}{(\lambda_{max}-\lambda_j)^2}+\ldots\right\}
\]
Using this expression one can try, in principle, to study statistics of the perturbed ground state by relating it to properties of random matrices.
For example, to the first order in $\sigma$ the ground state is equal to the sum of two independent variables since eigenvalues and eigenvectors of the random matrix are independent of each other. As is well-known, in the large-$N$ limit $\lambda_{max}=2J(1+{\bf \frac{1}{2}} \zeta N^{-2/3})$, with random $\zeta$ following the $\beta=1$ Tracy-Widom distribution \cite{TW}. On the other hand, it is easy to see that $w_j=w$ are all distributed with the probability density ${\cal P}(w)=\frac{1}{\sqrt{2\pi w}}e^{-w/2}$.
The ground state distribution is then the simple convolution of the two. Much less trivial are terms of the order $\sigma^2$ and higher in the series (\ref{per2}). In the language of the random matrix theory the second-order term can be interpreted as the so-called "level curvature'' associated with the largest eigenvalue. To find the distribution of this particular type of level curvature is a rather challenging RMT problem not yet solved (see a detailed discussion and description of the problematic for GUE matrices in \cite{Fyocurv12}), though in the bulk of the spectrum related objects for GOE were successfully investigated long ago \cite{Oppen95,Fyocurv95}.

One also can use the perturbation expansion (\ref{per2}) to estimate the scale $\sigma$ of the magnetic field at which all terms in the series for
$\delta{{\cal E}}_m=\frac{ E_{min}(h)-E_{min}^{(typ)}(h=0)}{NJ}$ become typically of the same order. Using that for $N\gg 1$ the typical eigenvalue separation between the $\lambda_{max}$ and the second largest eigenvalue is of the order $\Delta\sim J N^{-2/3}$ we see that the scale in question is given by $\sigma\sim \sqrt{N}\Delta\sim JN^{-1/6}$. For such values of the magnetic field we then have  $\delta{{\cal E}}_m\sim \Delta/J\sim N^{-2/3}$. It is natural to expect that in such a regime the probability density of the scaled random variable $\zeta=\delta{{\cal E}}_m N^{2/3}$ will be given by a {\it universal family} of distributions shared by minimization problems (\ref{1}) for  a broad class of random matrix ensembles and of the magnetic field distribution. The family is parametrized by the scaling variable $\kappa=N^{1/3}\sigma^2/J^2$ and is a very natural generalization of the Tracy-Widom law (and contains the latter as a limiting case at $\kappa=0$). To understand it properties is yet another challenging open problem. In the section 4 we will be able to understand far tail of such a distribution from matching to the large deviation result.

Very similarly one can develop perturbation theory for small $\sigma$ around any solution $t_j^{(0)}=\lambda_j$ of the secular equation (\ref{2})
for $\sigma=0$, with $\lambda_j$ in the bulk of the spectrum $(-2J,2J)$. It will be of the same type as (\ref{per1},\ref{per2}), but with $\lambda_{max}$ replaced by $\lambda_j$. In fact around each $t_j^{(0)}$ we will have two perturbative solutions $t_j^{(\pm)}$ different
by the sign in front of the perturbative terms. Using that the typical eigenvalue separation $\Delta\sim JN^{-1}$ in the bulk of spectrum
one can estimate that all the terms of the perturbation theory are of the same order for $\sigma\sim JN^{-1/2}$. This is the same scaling
as anticipated for the regime of gradual trivialization of the landscape topology. We are going to study the phenomenon of topology trivialization
quantitatively in the next section.

\section{Two-stage trivialization of the cost function landscape topology: quantitative considerations.}

Let us denote the mean of the total number of all stationary points  for a random field on a manifold as
 ${\cal N}_{tot}$. General framework for calculating that number for stationary Gaussian fields was developed in \cite{Fyo04} for unconstrained case and extended in \cite{Auf1,Auf2} to the case of spherically constrained isotropic fields pertinent to our problem. As the most convenient expressions for the mean total number of points with a given index in the spherically constrained case, see (\ref{totalnumberk}) and (\ref{totalnumber}) below, were not written down explicitly in \cite{Auf1,Auf2} we give below a brief derivation using equation (5.2) of \cite{Auf2} as the starting point (an {\it ab initio} derivation following a somewhat different route can be found in \cite{Fyolec}). It concerns the mean number $\mathbb{E}\{C^{(k)}_N(B)\}$ of critical (stationary) points with a given {\it index} (i.e. the number of positive eigenvalues of the Hessian) $k=0,1,2,\ldots N-1$ such that the values of the cost function $E_{h}({\bf x})$ restricted to the sphere $|{\bf x}|=\sqrt{N}$ at those critical points lie in a Borel set $B\in \mathbb{R}$. That object was shown to be given for all $N$ by:
\begin{equation}
\fl \mathbb{E}\{C^{(k)}_N(B)\}=C(N,\nu',\nu) \int_{B}\mathbb{E}_{GOE}\left\{ e^{\frac{N}{2}(\lambda_{k+1}^2-y^2)}e^{-\frac{2N\nu''}{2\alpha^2}\left(\lambda_{k+1}-\frac{\nu'y}{\sqrt{2\nu''}}\right)^2} \right\} \,dy
\end{equation}
where the expectation in the right-hand side goes over the random variable $\lambda_{k+1}$ which is distributed as the $k+1$-th lowest eigenvalue of the standard GOE random matrices with the variance chosen to satisfy $J^2=1/2$.
In the above formula $\nu'=\frac{d}{dx}\nu(y)|_{y=1}, \nu''=\frac{d^2}{dx^2}\nu(y)|_{y=1}$ and it is valid for a centered isotropic Gaussian field on the sphere with a  covariance function $\nu(y)$ defined by the identity:
$\mathbb{E}\left\{E_{h}({\bf x}_1)\,E_{h}({\bf x}_2)\right\}=N\nu\left(\frac{1}{N}{\bf x}_1^T{\bf x}_2\right)$. We denoted $\alpha^2=\nu''+\nu'-\nu'^2$. The factor ${C}(N,\nu',\nu)$ is given explicitly by ${C}(N,\nu',\nu)=2\left(\frac{2\nu''N}{\pi\nu'\alpha^2}\right)^{1/2}\frac{\nu'}{\sqrt{2\nu''}}\left(\frac{\nu''}{\nu'}\right)^{N/2}$
\footnote{Note that the value of $C(N,\nu',\nu)$ given in eq.(5.2) of \cite{Auf2} misses the factor $\frac{\nu'}{\sqrt{2\nu''}}$.}.
To count the totality of all stationary points with a given index $k$ irrespective of the values taken by the cost functions we set $B$ to conside with the real line $\mathbb{R}$.   We then can perform the (Gaussian) integral over the variable $y$ and get
 \begin{equation}\label{totalnumberk}
\mathbb{E}\{C^{(k)}_N(\mathbb{R})\}=2\left(\frac{2\nu'}{\nu'+\nu''}\right)^{1/2}\left(\frac{\nu''}{\nu'}\right)^{N/2} \mathbb{E}_{GOE}\left\{e^{\frac{N}{2} \frac{\nu'-\nu''}{\nu'+\nu''}\lambda_{k+1}^2} \right\}
\end{equation}

  Finally we can sum over all index values $k$ and exploit the identity $\sum_{k}F(\lambda_k)=N\int F(\lambda)\rho(\lambda)\,d\lambda$ where $\rho(\lambda)=\frac{1}{N}\sum_{k=0}^{N-1}\delta(\lambda-\lambda_{k+1})$ is the exact eigenvalue density of GOE. As the result we arrive at the following compact expression of general validity for the total number ${\cal N}_{tot}$ of stationary points in the spherical model:
\begin{equation}\label{totalnumber}
{\cal N}_{tot}=2N\left(\frac{2\nu'}{\nu'+\nu''}\right)^{1/2}\left(\frac{\nu''}{\nu'}\right)^{N/2}\int_{-\infty}^{\infty} \mathbb{E}_{GOE}\{\rho_N(\lambda)\}e^{\frac{N}{2} \frac{\nu'-\nu''}{\nu'+\nu''}\lambda^2} \,d\lambda
\end{equation}
In the so-called "pure" case $\nu(y)=y^p$, of a $p$-spin model, considered in \cite{Auf1} we have $\nu'=p,\nu''=p(p-1)$ and the eq.(\ref{totalnumber}) reproduces eq.(2.9)
of that paper.

It is easy to see that the cost function (\ref{1}) corresponds to the choice $\nu(y)=\frac{J^2}{2}y^2+\sigma^2 y$ which yields $\nu'=J^2 +\sigma^2, \,\nu''=J^2$ and eq. (\ref{totalnumber}) assumes the form
{\footnote{Note that though the treatment of \cite{Auf2} was formally restricted to covariances of the form $\nu(y)=y^2+\ldots $, one can check that inclusion of the linear term in the expansion does not invalidate their formalism.}:
\begin{equation}\label{totalnumber1}
\fl {\cal N}_{tot}=2N\left(\frac{2(J^2+\sigma^2)}{2J^2+\sigma^2}\right)^{1/2}\left(\frac{J^2}{J^2+\sigma^2}\right)^{N/2}
\int_{-\infty}^{\infty} \mathbb{E}_{GOE}\{\rho_N(\lambda)\}e^{\frac{N}{2} \frac{\sigma^2}{2J^2+\sigma^2}\lambda^2} \,d\lambda
\end{equation}
The above expression is exact for any $N$, and one can provide also the exact expression for the mean eigenvalue density $\mathbb{E}_{GOE}\{\rho_N(\lambda)\}$ in terms of the Hermite polynomials, see e.g. \cite{Mehta}
or eqs. (3.12)-(3.13) in \cite{For1}. As such it can be hopefully useful for comparison with the results of direct numerical simulations of the spherical model landscape, see \cite{MSK} for a recent work of that kind.
We however are interested mainly in the limit $N\to \infty$ where according to the earlier discussion we expect a nontrivial behaviour to occur at the scale $\sigma\sim N^{-1/2} J$. Indeed, introducing the parameter $\gamma=N\frac{\sigma^2}{2J^2}$ and performing the limit $N\to \infty$ for a fixed finite $\gamma$ we arrive at the following expression:
\begin{equation}\label{totalnumber2}
 \lim_{N\to \infty}\frac{{\cal N}_{tot}}{2N}= {\cal N}(\gamma)=e^{-\gamma}\int_{-\sqrt{2}}^{\sqrt{2}}\sqrt{2-\lambda^2}\,
e^{\frac{\gamma}{2} \lambda^2} \,\frac{d\lambda}{\pi}, \quad \gamma=N\frac{\sigma^2}{2J^2}
\end{equation}
where we have used that the limiting eigenvalue density has the semicircular profile {\bf (\ref{4})} in the interval $(-\sqrt{2},\sqrt{2})$. One can further simplify this expression by introducing $\lambda=\sqrt{2}\cos{\theta}, \theta \in[0,\pi]$ and noticing that the resulting integral can be related to the Bessel function of imaginary argument $I_0(z)$. This yields finally the expression
\begin{equation}\label{totalnumber3}
 {\cal N}(\gamma)=-2\frac{d}{d\gamma} \left(e^{-\frac{\gamma}{2}}I_0\left(\frac{\gamma}{2}\right)\right)
\end{equation}
 with the small-$\gamma$ expansion $ {\cal N}(\gamma)= 1 - \frac{3 \gamma}{4} + \frac{5}{16} \gamma^2 + O(\gamma^3)$.
   In particular, $ {\cal N}(\gamma=0)=1$ and monotonically decreases with growing $\gamma$, being of the order of unity for any finite $\gamma<\infty$. This function is plotted in Fig. \ref{fig:fig1}.

We conclude that for any $\sigma/J\sim N^{-1/2}$ the total number of stationary points is asymptotically of the order of $N$ and therefore one may  expect nontrivial aging effects to take place.
 Let us mention that it is natural to expect that the mechanism of reduction of the number of real solutions of the secular equation (\ref{2}) is by pairwise collisions of the real roots as a function of the growing magnetic field and disappearance of the pair into the complex plane. The last removed are to be stationary points corresponding to the Legendre multipliers $t$ with values close to the spectral edges $\pm 2J$. Analytical and numerical
 understanding of that picture, as well as investigating statistics of solutions of the equation (\ref{2}) in the crossover regime, and statistics of the cost function values (energies) at critical  points at finite $\gamma$ seem to us as interesting open problems deserving further attention.

\begin{figure}[htpb]
  \centering
 \includegraphics[width=.8\textwidth]{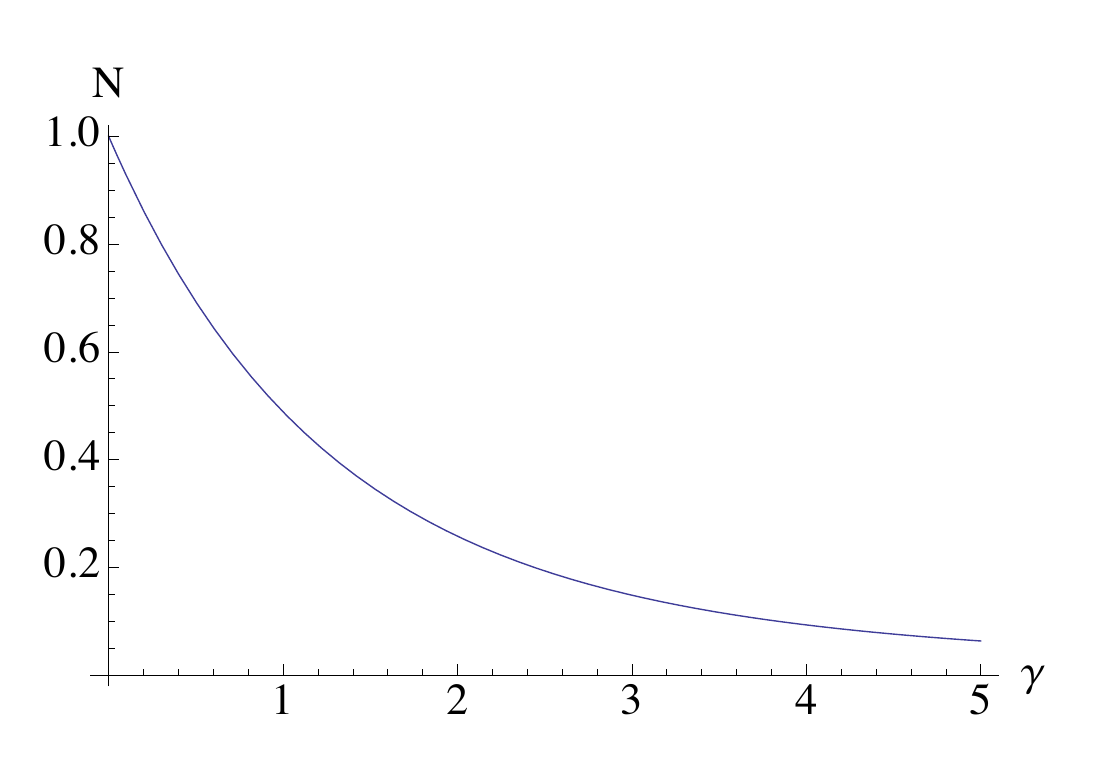}
  \caption{Mean number of stationary points (divided by  $2N$) as a function of $\gamma=N \sigma^2/(2 J^2)$ in the first scaling regime $\sigma/J = O(N^{-1/2})$, from the formula (\ref{totalnumber3}).}
\label{fig:fig1}
\end{figure}

  The formula (\ref{totalnumber3}) can be further used to infer the existence of yet another relevant magnetic field scale $\sigma$ such that the process of landscape trivialisation enters its final stage.
   This happens when ${\cal N}_{tot}$ drops to the values of order of unity.
  Exploiting the asymptotic $I_0(z\gg 1)\sim e^{z}/\sqrt{2\pi z}$ we obtain  $ {\cal N}(\gamma\gg 1 )\approx \frac{1}{\sqrt{\pi}}\gamma^{-3/2}$. Second stage then corresponds to  $ {\cal N}(\gamma)$ of the order of $1/N$ which occurs at $\gamma\sim N^{2/3}$, that is $\sigma/J\sim N^{-1/6}$. From our previous consideration we have seen that this was precisely the scale when the magnetic field term started to affect the statistics of the global energy minimum, with the scaling parameter now being $\kappa=2\gamma N^{-2/3}=N^{1/3}\sigma^2/J^2$.

  \begin{figure}[htpb]
  \centering
  \includegraphics[width=0.8\textwidth]{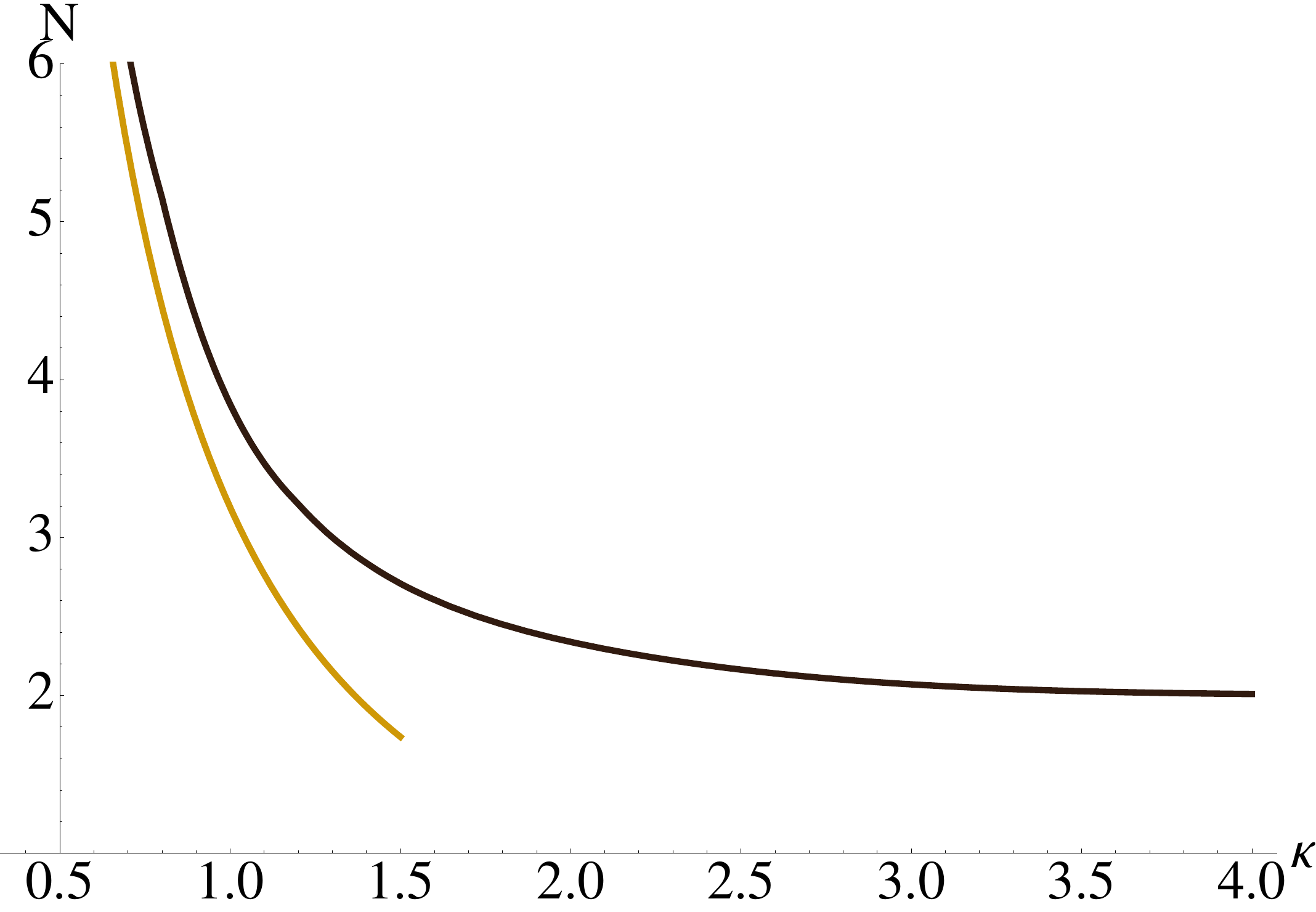}
  \caption{ Mean number of stationary points as a function of $\kappa=N^{1/3} \sigma^2/J^2$ in the second scaling regime $\sigma/J = O(N^{-1/6})$, from the formula (\ref{totalnumber3a}). The asymptotic
  formula for small $\kappa$, Eq. (\ref{totalnumber4}) is also indicated as the lower curve. For $\kappa \to \infty$ the mean number converges to the minimal possible value $2$ (see the text). }
\label{fig:fig2}
\end{figure}

 In fact, it is easy to understand that if we like to know precise number of stationary points in that new scaling regime the expression (\ref{totalnumber2}) should be replaced with a more accurate formula. This can be most easily seen by the fact that (\ref{totalnumber2}) in such a regime is dominated by the vicinities of the spectral edges $\lambda=\pm \sqrt{2}$ of the widths $|\lambda\pm \sqrt{2}|\sim N^{-2/3}$  where the semicircular law should be replaced by a more accurate expression. Using the symmetry $\mathbb{E}_{GOE}\{\rho_N(\lambda)\}=\mathbb{E}_{GOE}\{\rho_N(-\lambda)\}$ we can restrict integration in (\ref{totalnumber}) to $\lambda\in[0,\infty)$ multiplying the result by the factor of two, and so it is enough to consider the scaling vicinity of only one edge $\lambda=\sqrt{2}$. Introducing $\lambda=\sqrt{2}\left(1+\frac{\zeta}{2N^{2/3}}\right)$
 one finds that $\mathbb{E}_{GOE}\{\rho_N(\lambda)\approx N^{-1/3}\sqrt{2}\rho_{edge}(\zeta)$ where explicit expression for $\rho_{edge}(\zeta)$ is given by\cite{For1993}
 \begin{equation}\label{edgedens}
  \rho_{edge}(\zeta)=\left[Ai'(\zeta)\right]^2-\zeta\left[Ai(\zeta)\right]^2+\frac{1}{2}Ai(\zeta)
 \left(1-\int_{\zeta}^{\infty}
 Ai(\eta)\,d\eta\right)
\end{equation}
where $Ai(\zeta)=\frac{1}{2\pi i}\int_{\Gamma}^{}e^{\frac{v^3}{3}-v\zeta}$ is the Airy function solving the differential equation $Ai''(\zeta)-\zeta Ai(\zeta)=0$.

   Performing the corresponding limit $N\to \infty$  in (\ref{totalnumber1})  keeping $\kappa=N^{1/3}\sigma^2/J^2$ finite we get
 the exact expression for the limiting number of critical points in this regime as
 \begin{equation}\label{totalnumber3a}
  \lim_{N\to \infty}{\cal N}_{tot}={\cal N}_{tot}(\kappa)=4e^{-\kappa^3/24}\int_{-\infty}^{\infty} e^{\frac{\kappa}{2}\zeta}\rho_{edge}(\zeta)\,d\zeta, \quad \kappa=N^{1/3}\sigma^2/J^2
\end{equation}
 We see that it always remains of the order of unity. This function is plotted in Fig. \ref{fig:fig2}. We can easily extract the values for ${\cal N}_{tot}(\kappa)$ for $\kappa\ll 1$ and $\kappa\gg 1$ using the known asymptotic behaviour:
 \begin{equation}\label{edgedensa}
\fl  \rho_{edge}(\zeta\to -\infty)\approx \frac{\sqrt{|\zeta|}}{\pi} , \quad \rho_{edge}(\zeta\to +\infty)\approx \frac{1}{2} Ai(\zeta) \approx
   \frac{1}{4\sqrt{\pi} \zeta^{1/4}}\exp{\left\{-\frac{2}{3}\zeta^{3/2}\right\}}
\end{equation}
The small-$\kappa$ behaviour of ${\cal N}_{tot}(\kappa)$ is obviously controlled by $\zeta\to -\infty$ asymptotics, and we have:
\begin{equation}\label{totalnumber4}
 {\cal N}_{tot}(\kappa\ll 1)\approx 4\int_{-\infty}^{0}e^{\frac{\kappa}{2}\zeta}\, \frac{\sqrt{|\zeta|}}{\pi} d\zeta = \frac{4\sqrt{2}}{\sqrt{\pi} \kappa^{3/2}} \gg 1
\end{equation}
which precisely matches the $ {\cal N}(\gamma\gg 1 )\propto \gamma^{-3/2}$ behaviour obtained by us earlier. On the other hand, the behaviour ${\cal N}_{tot}(\kappa\gg 1)$ is controlled by $\zeta\to \infty$ asymptotic:
\begin{equation}\label{totalnumber5}
 \fl {\cal N}_{tot}(\kappa\gg 1)\approx \frac{e^{-\kappa^3/24}}{\sqrt{\pi}}\int_{0}^{\infty} e^{-\frac{2}{3}\zeta^{\frac{3}{2}}+\frac{\kappa}{2}\zeta}\, \frac{1}{\zeta^{1/4}} d\zeta =\frac{e^{-\kappa^3/24}\kappa^{3/2}} {\sqrt{\pi}}\int_{0}^{\infty} e^{-\kappa^3\left(\frac{2}{3}u^{\frac{3}{2}}-\frac{u}{2}\right)}\, \frac{du}{u^{1/4}}
\end{equation}
where we have made a substitution $\zeta=u\,\kappa^2$ to make it evident that the integral in the limit $\kappa\gg 1$ can be evaluated by the Laplace method around the stationary point $u=1/4$.  Equivalently we can use (\ref{edgedensa}) and the identity
\begin{equation} \label{AiryIdentity}
\int_{-\infty}^{+\infty} d\zeta Ai(\zeta) e^{\frac{\kappa}{2} \zeta} =e^{\frac{\kappa^3}{24}}
\end{equation}
for any $\kappa \geq 0$. The straightforward calculation
then yields $\lim_{\kappa\to \infty}{\cal N}_{tot}(\kappa\gg 1)=2$. This is the minimal possible value implying
the existence of a single minimum and single maximum only.

  In fact not only the mean total number of all critical points of the cost functional, but the mean number of true extrema  (minima or maxima) can be found in explicit form, and in the scaling regime $\sigma/J\sim N^{-1/6}$ is very directly related to the famous Tracy-Widom distribution \cite{TW}. Indeed, minima correspond to the index $k=0$, and their mean number is accounted by $\mathbb{E}\{C^{(k=0)}_N(\mathbb{R})\}$ from (\ref{totalnumberk}) so is related to the statistics of the minimum eigenvalue $\lambda_1$ (cf. \cite{FyNa12}). Introducing now
  the random variable $\zeta$ by $\lambda_{1}=\lambda_{min}=-\sqrt{2}\left(1+\frac{\zeta}{2N^{2/3}}\right)$ and performing the limit  $N\to \infty$ in  (\ref{totalnumberk}) keeping $\kappa$ finite we express the mean number of minima (or maxima) as:
\begin{equation}\label{totalnumberextrema}
 \fl \lim_{N\to \infty}\mathbb{E}\{C^{(k=0)}_N(\mathbb{R})\}= {\cal N}_m(\kappa)=2e^{-\kappa^3/24}\mathbb{E}_{\zeta}\left\{e^{\frac{\kappa}{2} \zeta}\right\}=2e^{-\kappa^3/24}
 \int_{-\infty}^{\infty} e^{\frac{\kappa}{2} \zeta}F'(\zeta)  d\zeta
\end{equation}
where $F'(\zeta)=\frac{dF_1}{d\zeta}$ and
\begin{equation}\label{TWdistr}
F_1(\zeta)=Prob\left\{\lambda_{max}\le \sqrt{2}\left(1+\frac{\zeta}{2N^{2/3}}\right)\right\}
\end{equation}
is the Tracy-Widom distribution \cite{TW}. By definition ${\cal N}_m(\kappa\to 0)=2[F_1(\infty)-F_1(-\infty)]=2$.
Near $\kappa=0$ one finds ${\cal N}_m(\kappa)= 2 -1.20652 \kappa + O(\kappa^2)$. For large $\kappa$ the integral is controlled by the
tail of the Tracy-Widom distribution, which takes the form:
\be
\frac{dF_1}{d\zeta}|_{\zeta\gg 1}\approx \rho_{edge}(\zeta \gg 1) \approx \frac{1}{2} Ai(\zeta \gg 1) \approx
\frac{1}{4\sqrt{\pi} \zeta^{1/4}}\exp{\left\{-\frac{2}{3}\zeta^{3/2}\right\}} \label{tailTW}
\ee}
Exploiting (\ref{totalnumber5}) or equivalently (\ref{AiryIdentity})  we find that $\lim_{\kappa\gg 1}{\cal N}_m(\kappa)=1$. The function
is plotted in Fig. \ref{fig:fig3}.

We thus see that in the second ("edge") scaling region $\sigma\sim N^{-1/6}$ the growing magnetic field gradually reduces the mean number of minima from two to just a single minimum. It is also easy to check that the mean number of minima always remains equal to two in the first ("bulk") scaling limit $\sigma\sim N^{-1/2}$, and we have already seen it is equal to one for any field of the order of unity. This corresponds to the following picture: initially at zero field among $2N$ critical points of the cost function there existed two global minima with exactly equal energies $E_{min}=-\frac{1}{2} N \lambda_{max}$ whose position vectors were related by the reflection ${\bf x}\to -{\bf x}$. Any nonzero magnetic field forces those two critical points to have slightly different energies but as  long as the magnitude $\sigma$ satisfies $\sigma/J\ll N^{-1/6}$ both of them with probability tending to unity retain their identity as minima. Only when $\sigma/J\sim N^{-1/6}$ the highest of the two minima has a nonvanishing probability to be converted to a saddle-point with nonzero index, the probability being higher the bigger is the value of $\kappa=N^{1/3}\sigma^2/J^2$. Finally, for $\kappa\to \infty$ (and in particular, for $\sigma/J\sim 1$) the probability of having only single minimum in the energy landscape tends to unity when $N\to \infty$.

  \begin{figure}[htpb]
  \centering
 \includegraphics[width=0.8\textwidth]{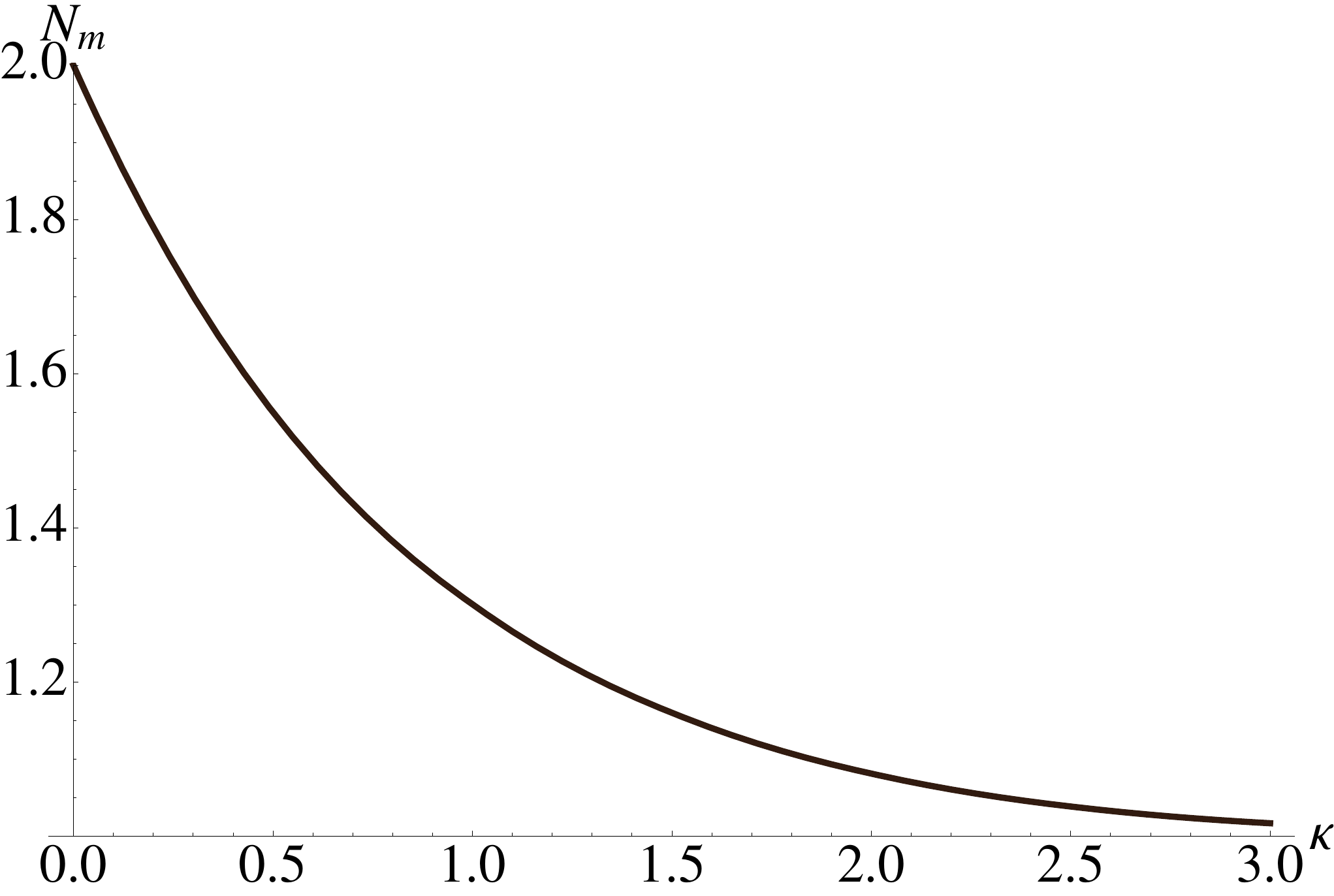}
  \caption{ Mean number of extrema as a function of $\kappa=N^{1/3} \sigma^2/J^2$ in the second regime
 $\sigma/J = O(N^{-1/6})$, from the formula (\ref{totalnumberextrema}). The number varies from $2$ at $\kappa=0$ to $1$ for $\kappa \to \infty$ }
\label{fig:fig3}
\end{figure}

Let us finally mention that the formula (\ref{totalnumberextrema}) is closely related to one derived in \cite{FyNa12} in the course of studies of the mean number of minima in a rather different model of high-dimensional random potential without spherical constraint. Namely, that type of model is known to display a zero-temperature transition to the phase with broken replica symmetry, and the Tracy-Widom distribution was shown to play a role in describing the change of the counting function of minima in  the scaling vicinity of the phase transition. This fact points towards a certain universality of our results. For a detailed discussion of this and related issues in a broader context see \cite{Fyolec}.

\section{Replica trick I: extracting the large deviations rate and pre-exponential factors from partition function moments.}

To employ the replica method  for our minimization problem we treat it as a problem of Statistical Mechanics, see e.g. \cite{Parisilec04}. Allowing for the temperature $T>0$ we start with introducing the partition function associated with the model
\be\label{6}
 {\cal Z}(\beta)=\int e^{-\beta E_h({\bf x})}\delta\left({\bf x}^{T}\,{\bf x}-N\right) d{\bf x}, \quad  d{\bf x}=\prod_{i=1}^N dx_i, \quad \beta=T^{-1}\,,
\ee
and consider the integer moments $\left\langle  {\cal Z}^n(\beta)\right\rangle $. The Gaussian nature of $E_h({\bf x})$ allows us to perform the ensemble average easily.
 In particular, rewriting $\sum_{a=1}^n{\bf x}_a^{T} H {\bf x}_a=\mbox{Tr}\left[H \sum_{a=1}^n{\bf x}_a \otimes {\bf x}_a^T\right]$ allows to perform the averaging over $H\in GOE$ by using the identity $\left\langle \exp{-\mbox{\small Tr}\left[HA\right]}\right\rangle =\exp{\frac{J^2}{4N}}\mbox{\small Tr}\left(A+A^T\right)^2$ valid for any matrix $A$. Similarly, $\left\langle \exp{\beta {\bf h}^T\sum_{a=1}^n{\bf x}_a}\right\rangle =\exp{\left(\frac{\beta^2\sigma^2}{2}\sum_{a,b}^n{\bf x}^T_a{\bf x}_b\right)}$. The specific rotational invariance of the integrand after the averaging is performed allows then at the next step to follow the method of \cite{FS}, see eqs. (10)-(11) of that paper.
 To that end one introduces the $n\times n$ positive semi-definite real symmetric matrix $Q$ of scalar products
with entries $q_{ab}=({\bf x}_a^{\dagger}{\bf x}_b)$ and uses $q_{a \leq b}$ as $n(n+1)/2$ new integration variables. Changing after that the scale $Q\to NQ$ we get
in the standard way
\bea \label{7}
&& \fl \left\langle{\cal Z}^n\right\rangle=C_{N,n}N^{-n}\int_{Q>0}
\,e^{N\frac{\beta^2}{2}\left[\frac{ J^2}{2} \mbox{\footnotesize Tr}Q^2+\sigma^2\sum_{a,b}q_{ab}\right]} \det{Q}^{(N-n-1)/2}  \prod_{a=1}^n
\delta\left(q_{aa}-1\right)  dQ \\
&& = C_{N,n}N^{-n}\int_{Q>0}
\,e^{N \Phi_n(Q)} \det{Q}^{(-n-1)/2}  \prod_{a=1}^n
\delta\left(q_{aa}-1\right)  dQ
\eea
where $ C_{N,n}=N^{nN/2}
\frac{\pi^{\frac{n}{2}\left(N-\frac{n-1}{2}\right)}}
{\prod_{k=0}^{n-1}\Gamma\left(\frac{N-k}{2}\right)}$  and we assumed $N\ge n+1$. In the large-$N$ limit the form of the integrand is suggestive of the saddle-point method with the functional to be extremized given by:
\begin{equation}\label{8}
\Phi_n(Q)=\frac{\beta^2J^2}{4} \mbox{Tr}(Q^2)+\frac{\beta^2\sigma^2}{2}\sum_{a,b}^nq_{ab}+\frac{1}{2}\mbox{Tr}\ln{Q}
\end{equation}
so that the stationarity conditions are
\begin{equation}\label{9}
\frac{ \partial}{\partial q_{ab}} \Phi_n(Q)=\beta^2J^2\, q_{ab}+\beta^2\sigma^2+ \left(Q^{-1}\right)_{ab}=0, \quad \forall a<b
\end{equation}
 Looking for the relevant saddle-point to be replica-symmetric: $q_{aa}=1, \, q_{a<b}=q, \quad \forall a<b$ we find that the inverse $Q^{-1}$ has a similar structure with diagonal/off-diagonal entries given by
 \begin{equation}\label{RS}
 \fl p_d=\left(Q^{-1}\right)_{aa}=\frac{1+q(n-2)}{(1-q)(1+q(n-1)}, \quad \quad p=\left(Q^{-1}\right)_{a\ne b}=-\frac{q}{(1-q)(1+q(n-1)}
 \end{equation}
  It is also easy to show that
  \be
  \det{Q}=(1+q(n-1))(1-q)^{n-1} \label{detQ}
  \ee
 Substituting the latter formula to (\ref{9}) we can bring it to the form
 \begin{equation}\label{10}
(J^2q+ \sigma^2) (1-q)\left(1+q(n-1)\right)-T^2q=0
\end{equation}
 Here we study this equation analytically continued for
$n=0$ and $n$ near zero. Then there are generically three roots. Excluding the solution with $q>1$ leaves two roots, e.g. for $\sigma=0$ these are $q=0,1-T$. For $T<1$ the root
$q=1-T$ is the physical solution corresponding to a (replica-symmetric) spin glass phase, which is essentially a "disguised ferromagnetic" \cite{DG}. For $T>1$ it is $q=0$ (paramagnetic phase).
In presence of a random field $\sigma>0$, the case studied here, the transition at $T=1$ disappears: one root lies
in the interval $0<q<1$ and is the physically relevant one, while the second root has $q<0$ and should not be considered. Here, in addition
we will be interested in the optimization problem, i.e. the zero $T$ limit.

 One has then $\left\langle{\cal Z}^n\right\rangle \sim e^{\Phi_n(Q)|_{sp}}$ where the
functional at the saddle point takes the value :
\bea
\fl && \frac{\Phi_n(Q)|_{sp}}{n} = \frac{\beta^2 J^2}{4}(1+ (n-1)q^2) + \frac{\beta^2 \sigma^2}{2}(1+ (n-1)q)\\
&& + \frac{1}{2 n} (\ln(1+q(n-1))+(n-1) \ln(1-q))
\eea

The standard use of the replica trick is for extracting the ensemble-averaged free energy per degree of freedom $ \langle f \rangle =-T\lim_{N\to \infty}N^{-1}\left\langle \ln{{\cal Z}(\beta)}\right\rangle$ which can be done in the replica formalism as
\bea \label{11}
\fl && - \langle f \rangle = \lim_{N\to \infty, n\to 0}T\frac{\ln \left\langle{\cal Z}^n\right\rangle}{Nn}=T\lim_{n\to 0} \frac{\Phi_n(Q)}{n}\\
&& =\frac{J^2}{4T}(1-q^2)+\frac{T}{2}\ln(1-q)+\frac{T}{2}\frac{q}{(1-q)}+\sigma^2\frac{1}{2T}(1-q)
\eea
where $q$ is the solution of (\ref{10}) for $n=0$. By definition, the zero-temperature limit of the mean free energy should coincide with the mean of the absolute minimum of the  energy functional {\it per degree of freedom} ${\sf e}_{min}=E_{min}(h)/N$, that is
$\lim_{T\to 0}\left\langle f \right\rangle=\langle {\sf e}_{min} \rangle$. Moreover, as $f$ is known to be self-averaging the mean and the typical value should coincide. Indeed, by solving (\ref{10}) in the limit $T\ll 1$ we easily find $q=1-Tv$ with $v=(J^2+\sigma^2)^{-1/2}$. Substituting this to (\ref{11}) and sending $T\to 0$ gives the finite value
\[
-\lim_{T\to 0}\left(T\lim_{n\to 0} \frac{\Phi_n(Q)}{n}\right)=-\frac{1}{2}\left[v(J^2+\sigma^2)+\frac{1}{v}\right]=-\sqrt{J^2+\sigma^2}
\]
 thus indeed reproducing the value $e^{(typ)}_{min}(\sigma) = E^{(typ)}_{min}(\sigma)/N$ from (\ref{5}).

One may however observe, cf. \cite{PR08}, that the low-temperature behaviour of the moments $\left\langle{\cal Z}^n\right\rangle$ can in fact be used not only for extracting  the mean $\langle E_{min}\rangle(\sigma)$, but rather obtaining the whole large deviation functional of the distribution of the random variable $E_{min}(h)$.
We start with assuming that the probability density ${\cal P}(E)$ of $E=E_{min}(h)$ takes in the thermodynamic limit $N\gg 1$ a well-defined large-deviation asymptotic form
\be \label{ratedef}
{\cal P}(E) \approx  R({\sf e})\, e^{- N {\cal L}({\sf e})} \quad , \quad {\sf e}=E/N
\ee
with the rate ${\cal L}({\sf e})$ and the leading pre-exponential factor $R({\sf e})$. On the
other hand we will see below that by scaling the replica index $n$ with temperature as $n=sT$ and keeping $s$ finite when both $T$ and $n$ tend to zero one can also define two functions $g(s)$ and $\phi(s)$ from
our (analytically continued) moments in the large $N$ limit:
\be
\lim_{n=s T, T\to 0}\left\langle{\cal Z}^n\right\rangle \approx  g(s) e^{N\phi(s)}
\ee
 Hence exploiting $ \ln{{\cal Z}(\beta)}=-\frac{N f}{T}$ and $\lim f|_{T\to 0}={\sf e}_{min}(h)$ we can now write the chain of identities:
\be \label{12}
\lim_{n=s T, T\to 0}\left\langle{\cal Z}^n\right\rangle=\lim_{T\to 0}\left\langle e^{-Ns f}\right\rangle=\left\langle e^{-N\, s\, {\sf e}_{min}(h)}\right\rangle
\ee
\be\label{12a}
\approx \int e^{-N\left(s {\sf e} +{\cal L}({\sf e})\right)}\, R({\sf e}) \, dE \approx  g(s) e^{N\phi(s)}
\ee
where in the last step we have applied the saddle-point method for evaluating the integral over the energy $E$ yielding the following
consistency relation between these functions:
\be \label{legendre}
\phi(s)=-\min_{{\sf e}}\left(s {\sf e}+{\cal L}({\sf e})\right)
\ee
We therefore see that $\phi(s)$ is the Legendre transform  of the large-deviation rate function ${\cal L}({\sf e})$. Moreover, the same procedure allows to relate the pre-exponential factor $R({\sf e})$ to $g(s)$ and $\phi(s)$. Namely, by recovering ${\cal P}(E)$ from  its  Laplace transform with help of the Bromwich integral and employing again the saddle-point method we find that asymptotically
\be \label{preexp}
\fl {\cal P}(E)\approx \int_{const-i\infty}^{const+i\infty} g(s) e^{N(s {\sf e}+\phi(s))}\,\frac{ds}{2i\pi}\approx
\frac{g(s_*)}{\sqrt{2 N \pi|\phi''(s_*)|}}  e^{N\left({\sf e}s_*+\phi(s_*)\right)}, \quad {\sf e}=-\phi'(s_*)
\ee

Now we proceed with implementing this program, first for finding the rate function, and then for the pre-exponential factors.

\subsection{Rate function calculation}

We make for small temperatures the Ansatz: $n=sT,\, q=1-vT$ where $v>0$ is expected to remain finite
when $T\to 0$.  The saddle-point equation (\ref{10}) in the limit of small $0<T\ll J^2$ takes the temperature-independent form $(J^2+\sigma^2)v (v+s)-1=0$ which is solved by
\be\label{13}
 v=\frac{1}{2}\left(-s+\sqrt{s^2+4B^2}\right), \quad B^2=\frac{1}{J^2+\sigma^2}
\ee
Similarly, the functional $\Phi_n(Q)$ from (\ref{8}) is transformed by the same low-temperature Ansatz to:
  \be\label{14}
\Phi(s,v)=\frac{J^2}{4}s(2v+s)+\frac{\sigma^2}{2}s(s+v)+\frac{1}{2}\ln{\left(1+\frac{s}{v}\right)}
\ee
which after substitution of the solution (\ref{13}) yields the Legendre transform $\phi(s)$ of the rate function in the final form:
  \be\label{15}
\phi(s)=\frac{\sigma^2}{4}s^2 + \frac{1}{4B^2}s\sqrt{s^2+4B^2}+\ln{\left(\frac{s+\sqrt{s^2+4B^2}}{2B}\right)}
\ee
The large deviation rate function ${\cal L}({\sf e})$ of the ground state energy can be found by ${\cal L}({\sf e})=- {\sf e}s_{*}-\phi(s_*)$  where
$s_*$ is the solution of $- {\sf e}=\phi'(s)=\frac{1}{2}\left(\sigma^2s+\frac{1}{B^2}\sqrt{s^2+4B^2}\right)$. For
any ${\sf e}<{\sf e}_c$, where ${\sf e}_c$ is the threshold:
\be\label{threshold}
{\sf e}_c = - J \sqrt{\frac{J^2 + 2 \sigma^2}{J^2 + \sigma^2}}
\ee
there are two roots to this equation:
\be \label{spm}
s_\pm^* = \frac{2}{J^2(J^2+2 \sigma^2)} ({\sf e} \sigma^2 \pm (J^2+\sigma^2) \sqrt{{\sf e}^2 - {\sf e}_c^2} )
\ee
which merge at ${\sf e}_c$.
One can check that only the $+$ root satisfies the requirement (\ref{legendre}) that the extremum is a minimum, hence we retain it. For
${\sf e} > {\sf e}_c$ there is no solution (we consider ${\sf e}<0$). In contrast to the $\sigma=0$ case note that the
typical (intensive) energy and the threshold are now distinct with ${\sf e}^{typ} < {\sf e}_c$.
\footnote{Note that $s^*$ vanishes at the typical energy and becomes negative for ${\sf e}_{typ} < {\sf e} < {\sf e}_c$, i.e. that region is controlled
by negative number of replica.}

Introducing the dimensionless variables  ${\cal E}={\sf e}/J=E/(N J)$ and $\Gamma=\sigma^2/J^2$ and denoting by the same letter
${\cal L}({\sf e}) \equiv {\cal L}({\cal E})$, we find after straightforward manipulations:
\bea
&& \fl {\cal L}({\cal E})=\frac{-{\cal E}}{1+2\Gamma} \left[{\cal E}\Gamma +(1+\Gamma)\sqrt{{\cal E}^2-\frac{1+2\Gamma}{1+\Gamma}}\right]
-\ln{\left\{\frac{\sqrt{1+\Gamma}}{1+2\Gamma}\left(- {\cal E}+\sqrt{{\cal E}^2-\frac{1+2\Gamma}{1+\Gamma}}\right)\right\}} \nonumber \\
&& \label{16}
\eea

\begin{figure}[htpb]
  \centering
  \includegraphics[width=0.8\textwidth]{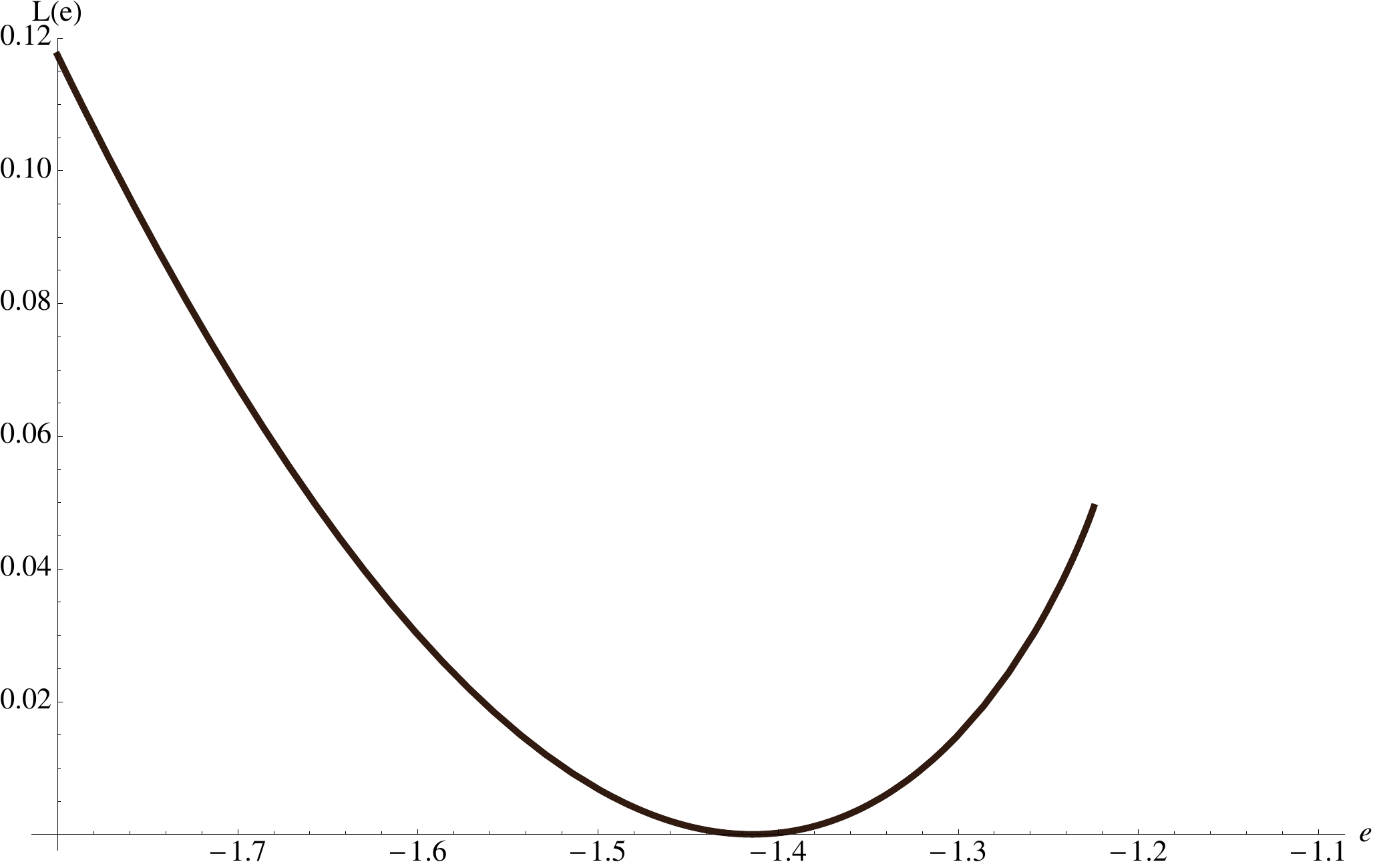}
  \caption{ Large deviation rate function ${\cal L}({\sf e})$ as a function of the (intensive) optimal energy ${\sf e}=E_{min}/N$, plotted for $\Gamma=1$, from Eq. (\ref{16}). The
  threshold is at ${\sf e}_c=\sqrt{3/2}=1.2247$ and the typical energy correspond to the minimum of the curve at ${\sf e}_{typ}=\sqrt{2}=1.41421$.}
\label{fig:fig5}
\end{figure}

This explicit formula for the large deviation rate function  ${\cal L}({\cal E})$ is one of the main results of the present paper.
Let us discuss the behavior of the rate function. It is defined only for ${\cal E}<{\cal E}_c=- \sqrt{\frac{1+ 2 \Gamma}{1+\Gamma}}$. Note that for ${\cal E}=-\sqrt{1+\Gamma}={\cal E}^{(typ)} < {\cal E}_c$ the rate function vanishes, and that value is simultaneously the minimum of ${\cal L}({\cal E})$ (see figure \ref{fig:fig5}).
This is consistent with the notion of ${\cal E}^{(typ)}$ as the typical value of the ground energy. Note also that in the limit of the vanishing magnetic field $\Gamma\to 0$ (\ref{16}) is reduced to
\be \label{L0}
{\cal L}({\cal E})={\cal L}_0({\cal E}) := - {\cal E}\sqrt{{\cal E}^2-1}-\ln{\left(- {\cal E}+\sqrt{{\cal E}^2-1}\right)}
\ee  which indeed coincides with the large deviation
rate function of the ${\cal E}=-\frac{1}{2J}\lambda_{max}$, with $\lambda_{max}$ being the maximal eigenvalue of GOE matrix \cite{BDG,MV}.
The present method does not say anything about large deviations for ${\cal E}^{(typ)}>{\cal E}_c$, but based on the RMT analogue \cite{DM,MV} one may conjecture that the rate function should in fact be infinite there, such that the probability of the ground state decaying as $\exp{(-N^2 const)}$  at $N\gg 1$, see also \cite{PR09,PR10}\footnote{The actual situation in the vicinity of ${\cal E}_c$ may appear to be even more complicated, see a note about the announced recent rigorous analysis of the problem by Dembo and Zeitouni in the Conclusion section.}.

 Around the typical value at fixed $\Gamma>0$ one has the following behaviour for ${\cal E} = {\cal E}^{(typ)} + y$:
\bea
&& {\cal L}({\cal E}) = \frac{y^2}{\Gamma }+\frac{(\Gamma +1)^{3/2} y^3}{3 \Gamma
   ^3}+O\left(y^4\right)
\eea
This implies the gaussian tails:
\be \label{gausstails}
P(E) \sim e^{- N ({\cal E}-{\cal E}^{(typ)})^2/\Gamma}, \quad J N^{1/2} \ll  |E- E^{(typ)}|  \ll NJ
\ee
It well may be that the distribution is exactly Gaussian in the regime of small deviations $E- E^{(typ)} \sim J N^{1/2}$, but formally our method
does not allow to infer the precise shape of the density in that regime. On the other hand, for vanishing magnetic field  $\Gamma=0$ we readily see $ {\cal L}({\cal E}) =  \frac{2}{3} (-2 y)^{3/2}$  which matches the exponent in the tail of the Tracy-Widom distribution (\ref{tailTW}) if we set $-2 y \equiv \zeta/N^{2/3}$.
\footnote{ for $\sigma=0$ the same calculation can be easily extended to any $T$ and the corresponding tail of the
free energy distribution $f$ (coming from the large deviation regime)
is found to be $\sim e^{- \frac{2}{3} (1-T)^3 (-2 y)^{3/2}}$ for $T<1$ with $f=f^{typ} + J y$, cf. \cite{PR09}. It would be interesting to investigate how the {\it small deviation} distribution of $f$ relates to the Tracy-Widom in the whole phase $T<1$}
In our language it can be seen as the consequence of $ {\sf e}_c={\sf e}^{typ}$ in this limit.
To this end it is worth to mention that for $\Gamma>0$ the $3/2$ power behaviour can be still seen in subleading terms of expansion around the threshold:
\bea
&& {\cal E} = {\cal E}_c - z \\
&& {\cal L}({\cal E}) =  \left(\frac{1}{2} \log (2 \Gamma +1)-\frac{\Gamma }{\Gamma
   +1}\right)-\frac{2 \Gamma  z}{\sqrt{(\Gamma +1) (2
   \Gamma +1)}}+O\left(z^{3/2}\right)
\eea
For small but finite $\Gamma\ll 1$ the large deviation function takes the following scaling behaviour
\be\label{smalltail}
\! \! \! \! \! \! \! \! \! \!  \! \! \! \! \! \! \! {\cal L}({\cal E}) = \Gamma^3 F(\frac{{\cal E} - {\cal E}^{typ}}{\Gamma^2}), \quad , \quad
 F(x) =  \frac{2}{3}  (\sqrt{1 - 2 x} -1 + (3 - 2 \sqrt{1 - 2 x}) x)
\ee
 where the function $F(x)$ is defined for $x \in ]-\infty,\frac{1}{2}]$.

As was discussed in the first chapter the results of the perturbation theory suggest that in the regime $\Gamma \sim  N^{-1/3}$
the probability of small deviations in the minimum energy from its typical value are expected to be given by a universal family of functions.
Using (\ref{smalltail}) we therefore can predict the tail behaviour for the densities belonging to that family. Namely, we expect that for $N\to \infty$ and
$ {\cal E} = {\cal E}^{typ} + \delta N^{- 2/3}, \quad \Gamma = \kappa N^{-1/3}$ the probability density ${\cal P}({{\cal E}})$ tends to  the function $p_{\kappa}(\delta)$ such that its tail for large negative $\delta$ and large positive $\kappa\gg 1$ has the form
\be\label{largetail}
 p_{\kappa}(\delta) \propto  e^{- \kappa^3 F(x)}, \quad \frac{\delta}{\kappa^2}=x<\infty
\ee
 where the scaling function $F(x)$, defined in (\ref{smalltail}), is {\it universal}. As mentioned above, for
$x \to -\infty$ one has $F(x) \sim \frac{2}{3} (-2 x)^{3/2}$ so as to match with the tail of the Tracy-Widom distribution (\ref{tailTW}) for $\sigma=0$ (with $\delta = - \zeta/2$). See also the formula (\ref{newinsertion})
below for the correct prefactor.

\subsection{Calculation of the pre-exponential factors}

To extract the leading pre-exponential factor in the present formalism we obviously must take into account
the Gaussian fluctuations around the replica-symmetric saddle-point solution. It is appropriate to mention that similar in spirit calculations were performed for extracting the subleading corrections to the mean minimum value of the cost functional in some random optimization problems, see e.g. \cite{flucdet1,flucdet2}.

To that end we start with combining (\ref{12}) and (\ref{7}) to write
\be \label{pe1}
\fl \left\langle e^{-Ns\, e_{min}}\right\rangle=\lim_{n=sT,T\to 0}\left\langle{\cal Z}^n\right\rangle\propto
e^{N\phi(s)} \lim_{n=sT,T\to 0}\det{Q}^{(-n-1)/2}\frac{1}{\sqrt{\det{\hat{A}}}}
\ee since the factor $C_{N,n} N^{-n} \approx 1$ in that limit \footnote{we use that
$\prod_{k=0}^{n-1} \Gamma((n-k)/2) = G(\frac{N+1}{2}) G(1+ \frac{N}{2})/(G(\frac{N-n+1}{2}) G(1+ \frac{N-n}{2}))$ in terms of the Barnes function $G(x)$}. Here $\hat{A}$ is the $n(n-1)/2$ matrix of the quadratic form describing the fluctuations around the saddle point whose entries are given by
\be\label{fl1}
\fl \hat{A}_{(ab)(cd)}=\frac{\partial^2\Phi_n(Q)}{\partial q_{(ab)}\partial q_{(cd)}}=\beta^2J^2\delta_{(ab),(cd)}-\left(Q^{-1}\right)_{(ac)}
\left(Q^{-1}\right)_{(bd)}-\left(Q^{-1}\right)_{(ad)}\left(Q^{-1}\right)_{(bc)}
\ee
where $a\ne b, c\ne d$. For the replica-symmetric saddle-point (\ref{RS}) the matrix $\hat{A}$ has three distinct elements:
\be\label{fl2}
\fl \hat{A}_{(ab)(ab)}=\beta^2J^2-(p_d^2+p^2)=A_1, \quad \hat{A}_{(ab)(ac)}=-(p_dp+p^2)=A_2,
\quad \hat{A}_{ (ab)(cd)}=-2p^2=A_3
\ee
The matrix of such structure was originally diagonalized in the course of the classical De-Almeida-Thouless stability analysis \cite{AT},
  revealing the existence of three distinct eigenvalues $\lambda_1,\lambda_2,\lambda_3$  given explicitly by
 \bea\label{fl3}
&& \!\!\!\!\!\!\!\!\!\!\!\!\!\!\!\!\!\! \!\!\!\!\!\!\!\!\!\!\!\!\!\ \lambda_1=A_1+2(n-2)A_2+\frac{(n-3)(n-2)}{2}A_3 \quad , \quad  \lambda_2=A_1+(n-4)A_2 - (n-3)A_3, \\
&& \!\!\!\!\!\!\!\!\!\!\!\!\!\!\!\!\!\! \!\!\!\!\!\!\!\!\!\!\!\!\!\ \lambda_{3}=A_1-2A_2+A_3
\eea
and the corresponding degeneracies are given by $d_1=1, d_2=n-1, d_3=\frac{n(n-3)}{2}$. We therefore see that
\[
\lim_{n=sT,T\to 0}\frac{1}{\sqrt{\det{\hat{A}}}}=\lim_{n=sT,T\to 0}\left(\lambda_1\lambda_2^{n-1}\lambda_3^{\frac{n(n-3)}{2}}\right)^{-1/2}=
\lim_{n=sT,T\to 0}\sqrt{\frac{\lambda_2}{\lambda_1}}
\]
Substituting here (\ref{fl3}),(\ref{fl2}) and (\ref{RS}) and further exploiting the low-temperature Ansatz $q=1- vT$  we find after straightforward calculations the low-temperature behaviour
 \be\label{fl4}
 \lambda_1 = -\frac{1}{T^3}\frac{2v+s}{v^2(v+s)^2}+O\left(\frac{1}{T^2}\right), \quad  \lambda_2 = -\frac{1}{T^3}\frac{2v+2s}{v^2(v+s)^2}+O\left(\frac{1}{T^2}\right)
\ee
which implies
\be\label{fl5}
 \left(\frac{\lambda_2}{\lambda_1}\right)^{1/2}_{T\to 0}=\sqrt{\frac{2v+2s}{2v+s}}
\ee
Similarly we have for the replica-symmetric $Q-$ matrices using (\ref{detQ})
\be\label{fl6}
 \left(\det{Q}^{(-n-1)/2}\right)_{n=sT, T\to 0}=\sqrt{\frac{v}{v+s}}
\ee
Combining all the factors together and using (\ref{13}) we finally arrive at the full asymptotic large-deviation expression for the Laplace transform of the probability density for the minimum:
\be \label{fl7}
\fl \left\langle e^{-Ns E_{min}(\sigma)}\right\rangle \approx g(s)\,e^{N\phi(s)}, \quad g(s)= \frac{2B}{\left(\sqrt{s^2+4B^2}(s+\sqrt{s^2+4B^2})\right)^{1/2}}
\ee
Note that $g(0)=1$ as required by normalisation. Now we can use (\ref{fl7}) and (\ref{preexp}) to recover the pre-exponential factor in the probability density ${\cal P}(E)$. Recalling the relation $\phi'(s)=\frac{1}{2}\left(\sigma^2s+\frac{1}{B^2}\sqrt{s^2+4B^2}\right)$ we first find
\be \label{fl8}
\phi''(s)= \frac{\left(s+\sigma^2B^2\sqrt{s^2+4B^2}\right)}{2B^2\sqrt{s^2+4B^2}}
\ee
and then using the relation between $s_*$  and ${\sf e}$ (\ref{spm})  we further establish the identities:
\be \label{fl9}
\fl s_*+\sigma^2B^2\sqrt{s_*^2+4B^2}=2B^2\sqrt{{\sf e}^2-{\sf e}_c^2}, \quad s_*+\sqrt{s_*^2+4B^2}=\frac{2}{J^2+2\sigma^2}\left(-{\sf e}+\sqrt{{\sf e}^2-{\sf e}_c^2}\right)
\ee
where the threshold $E_c=N {\sf e}_c$  was defined in (\ref{threshold}). Combining all the formulas we arrive at our final asymptotic large-deviation result
for the distribution of the minimum:
\be \label{fl20}
\fl {\cal P}(E)\approx \left(\frac{E_c^2}{ N \pi J^2\sqrt{E^2-E_c^2}(-E+\sqrt{E^2-E_c^2})}\right)^{1/2}e^{N{\cal L}({\sf e}=E/N)}, \quad  \quad E<E_c
\ee
 which is one of the main results of our paper.

 Several comments are in order. First for any $\Gamma>0$ one can expand this formula for ${\cal P}(E)$ around the most probable value (\ref{5}), i.e.
${\cal E}$ around ${\cal E}_{typ}=-\sqrt{1+\Gamma}$ as in (\ref{gausstails}) and obtain:
\be \label{norm}
{\cal P}(E) dE \equiv P({\cal E}) d {\cal E} \approx (\frac{N}{\Gamma \pi})^{1/2} e^{- \frac{N}{\Gamma} ({\cal E}-{\cal E}_{typ})^2} d {\cal E}
\ee
hence for any $\Gamma>0$ thanks to the prefactor it now reduces to a correctly normalized Gaussian distribution $\int {\cal P}(E) dE=1$ in the regime of
typical fluctuations.

Next if we naively take the limit $\Gamma=0$ of (\ref{fl20}) we find, for ${\cal E} <-1$:
\be \label{limit}
\lim_{\Gamma \to 0} P({\cal E}) d {\cal E} \approx \sqrt{\frac{N}{\pi}}
\frac{e^{- N {\cal L}_0({\cal E})} d {\cal E}}{({\cal E}^2-1)^{1/4} \sqrt{ - {\cal E} + \sqrt{{\cal E}^2-1}}}
\ee
Interestingly, the pre-exponential factor in (\ref{fl20}) has precisely the same structure
as the corresponding factor known from the independent non-trivial RMT calculations \cite{BorotNadal11,For1}.
If we compare (for convenience) with the formula (16) of Ref. \cite{FyNa12}, the variable denoted $s$ there
being $s  \equiv - \sqrt{2} {\cal E}$ (using again the choice $J^2=1/2$), we find that the limit
(\ref{limit}) is {\it exactly twice} the result (16) of Ref. \cite{FyNa12} \footnote{of course, as noted above
the exponent term is correct, i.e. ${\cal L}_0({\cal E}) = \psi_{+}(s)$ there.}. Similarly we can check the tail, replacing in (\ref{fl20})
$E \to - \frac{1}{2} N \lambda_{\max} = - \frac{N}{\sqrt{2}} (1+ \frac{\zeta}{2 N^{2/3}})$
one finds to leading order in large $N$:
\be
{\cal P}(E) dE \approx (\frac{2 N^{1/3}}{N \pi \sqrt{\zeta}})^{1/2} e^{ - \frac{2}{3} \zeta^{3/2}} dE=
\frac{1}{2 \sqrt{ \pi } \zeta^{1/4}} e^{ - \frac{2}{3} \zeta^{3/2}} d\zeta, \label{TWcheck}
\ee
which is also exactly twice the tail formula (\ref{tailTW}) for the TW distribution (which verifies
that the prefactor in (16) of Ref. \cite{FyNa12} matches exactly the large argument
limit of the TW law).

This mismatch of an overall factor of $2$ is puzzling at first, since we claim that a constant multiplicative factor could have been hardly missed in the calculation given the normalization property (\ref{norm}) noted above. After some thought one realizes that it is fixed $\Gamma>0$ which makes the above
saddle-point fluctuation calculation fully controlled at large $N$. The subtlety then likely arises due to a non-commutativity of the limits $\Gamma\to 0$ and $N\to \infty$ when the density ${\cal P}(E)$ ceases to be Gaussian in the vicinity of the most probable value. In that limit, i.e. strictly zero field $\Gamma=0$ first, the procedure (\ref{12a}) of inferring the pre-exponential factors in $ {\cal P}(E)$ from its Laplace transform in the large-N limit should be reexamined, as it was based on assuming the analyticity of the function $s {\sf e}+{\cal L}({\sf e})$ at the point of its minimum. A plausible scenario behind such a mismatch could be as follows. We have argued before that in the scaling regime $\Gamma\sim N^{-1/3}$ the probability density of the minimal energy in the small-deviation regime is given by a (presumably) universal family of densities parametrized by $\kappa=\Gamma N^{1/3}$, with the standard TW density recovered in the limit $\kappa=0$. If densities in the family contained a $\kappa-$ dependent multiplicative factor which changed smoothly between the values $1/4$ for $\kappa=0$ and $1/2$ for $\kappa\to \infty$ (cf. the behaviour of
the mean number of extrema in the same regime, Fig. 3), the limits $\Gamma\to 0$ and $N\to \infty$ would not commute in precisely the manner discussed above, explaining the observed mismatch.

Note that the factors in the exponentials match perfectly well, hence this is only a subtlety involving the fluctuations around the saddle point.
It is quite possible that the factor of $2$ could, in the end, be accounted by a one-sided only saddle point integration, but the details
are interesting and deserve to be further studied.

 Finally, it is also useful to reconsider the matching towards the small deviation regime from a slightly different perspective. As before we set $E= E_{typ} + N^{1/3} J \delta$, with both
 $\delta$  and $\kappa$ kept of order unity, i.e. $O(N^0)$, but eventually considered to be large. In that limit the two roots $s^*$ in (\ref{spm}) become
very close, hence one cannot rely on the Gaussian saddle point integration approximation (\ref{preexp}). Instead one
must recalculate more carefully the inverse Laplace transform from the formula:
\bea
&& {\cal P}(E)\approx \int_{const-i\infty}^{const+i\infty} g(s) e^{s E+N \phi(s)}\,\frac{ds}{2i\pi}
\eea
Introducing $s=N^{-1/3} \tilde s$ we find by expanding formula (\ref{15}) for $\phi(s)$ to cubic order in $s$:
\bea
s E+N \phi(s) = J \delta \tilde s + \frac{1}{4} J^2 \kappa \tilde s^2 + \frac{1}{24} J^3 \tilde s^3 + O(N^{-1/3})
\eea
 where we have used $E_{typ}=-N\sqrt{J^2+\sigma^2}\equiv-N/B$. Redefining $\tilde s= i z/J$, we find the density of the distribution of the variable $\delta$ (for a fixed $\kappa$) in the large $N$ limit to be given by:
\be\label{newinsertion}
 p_{\kappa}(\delta) \approx \int_{-\infty}^{+\infty} \frac{dz}{2 \pi} e^{-i \delta z - \frac{\kappa}{4} z^2 - \frac{i}{24} z^3}
 = 2 Ai(-2 \delta + \kappa^2)\, e^{- 2 \kappa \delta + \frac{2}{3} \kappa^3}
\ee
Although the right-hand side is normalized to unity on the whole real axis for $\delta$ this formula is expected to
be accurate only when both $-\delta$ and $\kappa$ are large. If one sets $\kappa=0$ in (\ref{newinsertion}) it again overestimates the asymptotics of the Tracy-Widom density by a factor 2. If one keeps in (\ref{newinsertion}) only the exponential asymptotics of the Airy function, $Ai(z) \sim e^{- \frac{2}{3} z^{3/2}}$ it
reproduces exactly the asymptotics (\ref{largetail}) in terms of the universal function $F(x)$ obtained in formula (\ref{smalltail}). However, we believe that (\ref{newinsertion}) does contain a bit more
information since it now displays the complete correct pre-exponential asymptotic factor.

\section{Replica trick II: direct approach to the distribution of the ground state.}

Let us now present an alternative way to extract the probability density of the minimum energy ${\cal E}=E_{min}(h)/NJ$ based on the identity:
\be \label{def1}
{\cal P}({\cal E}) = \lim_{\beta \to \infty} \overline{P_\beta({\cal E})}
\ee
where we have introduced
\bea
P_\beta({\cal E}) =  \left< \delta( {\cal E} - \frac{E_h(x)}{N J} ) \right>_\beta = \int _{-\infty}^{+\infty} \frac{dk}{2 \pi} e^{i k {\cal E}}  \left< e^{- i k \frac{E_h(x)}{N J} } \right>_\beta\,\,
\eea
with $\delta(u)$ being the Dirac delta-function and
\begin{equation}\label{Gibbs}
\left< .. \right>_\beta = \frac{1}{Z_\beta} \int d{\bf x} e^{\beta E_h(x)}
\end{equation}
 standing for the thermal average performed with the Gibbs measure for a single given realization of the disorder.
We can now use replica to express the disorder averages:
\bea
\overline{P_\beta({\cal E})}  = \lim_{n \to 0} \int _{-\infty}^{+\infty} \frac{dk}{2 \pi} e^{i k {\cal E}}  \overline{Z_\beta^{n-1} \int d{\bf x} e^{- (\beta + \frac{i k}{N J}) E_h(x)} }
\eea
We apply the same steps as before, the only difference being that one particular replica, labelled as $1$, is different from the rest of $n-1$ ones, leading to:
\be \label{def2}
\overline{P_\beta({\cal E})}  = C_{N,n}\int_{Q>0}
 \det{Q}^{(-n-1)/2}  \prod_{a=1}^n
\delta\left(q_{aa}-1\right)  dQ   e^{N \Psi_n(Q)}
\ee
with the new functional:
\bea
e^{N \Psi_n(Q)} : = \int _{-\infty}^{+\infty} \frac{dk}{2 \pi} e^{i k {\cal E}}  e^{N \Psi_n(Q,k)}
\eea
where we have defined:
\bea
N \Psi_n(Q,k) = N \Phi_n(Q) - \frac{k^2}{4 N} \frac{J^2 + 2 \sigma^2}{J^2} + i k \frac{\beta}{J} \sum_{a=1}^n (\frac{J^2}{2} q_{1a}^2 + \sigma^2 q_{1a})
\eea
using that $q_{11}=1$. Since the dependence in $k$ is quadratic we can perform the Gaussian integral over $k$ leading to our new functional:
\bea
\Psi_n(Q) = \Phi_n(Q)
- \frac{J^2}{J^2 + 2 \sigma^2} \big( {\cal E} + \frac{\beta}{J} \sum_{a=1}^n (\frac{J^2}{2} q_{1a}^2 + \sigma^2 q_{1a}) \big)^2
\eea
The saddle point equations read:
\bea
\fl &&  \left(Q^{-1}\right)_{ab} + \beta^2J^2\, q_{ab}+\beta^2\sigma^2 =0, \quad  1 < a<b \\
\fl && \left(Q^{-1}\right)_{1b} + \beta^2J^2\, q_{1b}+\beta^2\sigma^2 - \frac{2 \beta J}{J^2 + 2 \sigma^2} \big( {\cal E} + \frac{\beta}{J} \sum_{a=1}^n (\frac{J^2}{2} q_{1a}^2 + \sigma^2 q_{1a}) \big) (J^2 q_{1b} + \sigma^2)  \quad  b=2,..n \nonumber
\eea
It is natural to look for a replica symmetric solution with the following structure, $q_{aa}=1$, $q_{1b}=q_{b1}=u$, $b=2,..n$ and $q_{ab}=q_{ba} = q$ for $1<a<b$. Introducing
the inverse matrix with parameters $Q^{-1}_{11}=p_0$, $Q^{-1}_{aa}=p_d$ for $a\geq 2$, $Q^{-1}_{1b}=Q^{-1}_{b1} = \tilde u$ for $b\geq 2$,
$Q^{-1}_{ab}=Q^{-1}_{ba} = p$ for $b>a\geq 2$, we obtain the four equations:
\bea
&& p_0 + (n-1) u \tilde u = 1 \quad , \quad \tilde u + u(p_d+(n-2) p) = 0 \\
&& u p_0 + \tilde u (1+ (n-2) q) = 0 \quad , \quad u \tilde u + p_d + (n-2) q p = 1
\eea
Leading to:
\bea
\fl && p = \frac{u^2 - q}{(1-q)(1+ q(n-2)- (n-1) u^2)} \quad , \quad \tilde u =  \frac{-u}{1+ q(n-2)- (n-1) u^2} \\
\fl && p_0=\frac{1 + q(n-2) }{1+ q(n-2)- (n-1) u^2}  \quad , \quad p_d = \frac{1+ (n-3)q - (n-2) u^2}{(1-q)(1+ q(n-2)- (n-1) u^2}
\eea
This leads to the following saddle point equations in the limit $n=0$:
\bea
\fl && \frac{(u^2-q)T^2}{(1-q) (1-2 q + u^2)} + J^2 q + \sigma^2 = 0 \\
\fl && - \frac{T^2 u}{1- 2 q + u^2} + J^2 u + \sigma^2 = \frac{2 J (J^2 u + \sigma^2)}{J^2 + 2 \sigma^2} \big( {\cal E} T + \frac{1}{J} (\frac{J^2}{2} (1-u^2) + \sigma^2 (1-u))
\eea
We can solve these equations at low $T$ inserting the following expansion:
\bea
q= 1 - T v + T^2 w + O(T^3) \quad , \quad u= 1 - T v + T^2 r + O(T^3)
\eea
and we find   $v^2 + 2 {\cal E} v = - {\cal E}_c^2$
\bea
&&  v= - {\cal E} \pm \sqrt{{\cal E}^2 - {\cal E}_c^2} \quad , \quad {\cal E}_c = - \sqrt{\frac{1+2 \Gamma}{1+\Gamma} } \label{eqv} \\
&& r - w = 1 + {\cal E} v
\eea
 and we recall that $\Gamma=\sigma^2/J^2$.

To calculate the functional at the saddle point we need to evaluate the $\mbox{Tr}\ln{(Q)}$.
The eigenvalues of $Q$ are  displayed in the Appendix  of \cite{LDW}:
\bea
&& 1- q \quad , \quad \, \mbox{with multiplicity}\quad  d=n-2 \\
&& \mu_\pm = \frac{1}{2} (2 + (n-2)q \pm \sqrt{ (n-2)^2 q^2 + 4 u^2 (n-1)}
\eea
This leads to:
\bea
\det{Q} = (1-q)^{n-2} (1 + q (n-2) - u^2 (n-1))
\eea
and also for $n=0$:
\bea
&& \mbox{Tr} Q^2 = 2 (q^2 - u^2) \\
&& \mbox{Tr}\ln{(Q)} = - 2 \ln(1-q) + \ln(1- 2 q + u^2) \\
&& \sum_{ab} q_{ab} = 2(q-u)
\eea
which then gives:
\bea
&& \Psi_n(Q) = - \ln(1-q) + \frac{1}{2} \ln(1- 2 q + u^2)
+ \frac{\beta^2 J^2}{2} (q^2 - u^2) + \beta^2 \sigma^2 (q-u)  \\
&& - \frac{J^2}{J^2 + 2 \sigma^2} \big( {\cal E}  + \frac{\beta}{J} (\frac{J^2}{2} (1-u^2) + \sigma^2 (1-u)) \big)^2
\eea
Its zero temperature limit $T=0$ is found to be:
\bea
\lim_{T \to 0}  \Psi_n(Q) = (1+\Gamma) (w-r) - \frac{({\cal E} + (1+ \Gamma) v)^2 }{1+ 2 \Gamma} + \frac{1}{2} \ln( \frac{2 (r-w)+v^2}{v^2})
\eea
 Choosing the $-$ branch in (\ref{eqv}) we recover the formula (\ref{16}). More precisely, from (\ref{def1}),(\ref{def2}) and the
definition of the large deviation rate function (\ref{ratedef}) :
\be
{\cal P}(E) \sim e^{N \Psi_N(Q)} \quad , \quad {\cal L}(E) = - \lim_{T \to 0} \Psi_N(Q)
\ee
Hence this more direct method
to calculate the probability distribution gives an identical result to the more conventional method
of the previous Section using the analytical continuation from integer moments via the replica saddle point. While the previous method
used the scaling $n = s T$ the present method works directly at $n=0$.

\section{Conclusions and Open Problems}

We have demonstrated that despite its deceptive simplicity the problem of describing statistics of the minima of a cost function given by the sum of a random quadratic and a random linear form in $N$ real variables over $(N-1)-$ dimensional sphere has rather rich phenomenology, and generates quite a few open questions.
The existence of two nontrivial scaling regimes is intimately
connected with properties of random matrix spectra, and in a separate publication it will be demonstrated that essentially the same scenario of the topology trivialization takes place in a general spherical spinglass model with $p-$spin interaction in the scaling vicinity of the replica symmetry breaking point\cite{Fyolec}.

 Yet the standard RMT spectral methods and techniques, being  very useful for the problem of counting various types of critical points in the cost function landscape, do not seem to be of obvious utility for extracting the statistics of minima beyond the perturbation theory. Thus, for getting explicit analytical insights into the statistical characteristics of the global minimum we had to resort to the powerful heuristic method of Statistical Mechanics, the replica trick. Note that the replica methods have recently allowed to unveil the convergence to Tracy Widom distributions of the free energy of directed polymers in random media and of the height field of the Kardar-Parisi-Zhang growth equation \cite{KPZreplica1,KPZreplica2,KPZreplica3,KPZreplica4}, and it seems as an important goal to understand whether these approaches can extend to random matrices as well.
We have indeed found that the large-deviation results extending those known in the random matrix theory can be successfully reproduced by replica. To that end we should mention that our paper motivated Dembo and Zeitouni to perform a rigorous large-deviation analysis of the problem. Their method confirmed our formula (\ref{16}) in a certain range of the parameter ${\cal E}<{\cal E}_*$, where the value ${\cal E}_*$ lies in between the typical ${\cal E}_t$ and the threshold ${\cal E}_c$. Beyond that range the large deviation functional seems to be given by a different expression. To find a mechanism responsible for that change within our replica approach remains an interesting challenge, along with  extending these considerations to the level of (Tracy-Widom like) small deviations in the corresponding scaling regime as well as to investigating the issue of universality.

Even at the level of perturbation theory the problem touches on poorly explored RMT problems like parametric motion of extreme eigenvalues. In general, clarifying the RMT content of the quadratic eigenvalue problem in question, such as the gradual reduction of number of real solutions of the characteristic equation (\ref{2}), remains an interesting open task. It goes without saying that all the same questions can be asked, (and to the extent covered in the paper, answered) for complex quadratic and linear forms, with GUE matrices $H$ replacing the GOE ones.
Completely open is the question of investigating all aspects of the same problem for quadratic forms based on non-invariant ensembles of random matrices, such as various matrices with i.i.d. entries (Wigner, sparse, banded, etc.).

Finally, it is natural to expect that the zero-temperature gradient descent dynamics (or, more generally, Langevin dynamics with a noise simulating finite temperatures) should also reflect the existence of the two scaling regimes of the small magnetic field revealed by our considerations.

\section{Acknowledgements}
  We are grateful to Antonio Auffinger for a useful communication related to the content of the paper \cite{Auf2}, to Jean-Philippe Bouchaud and  Satya Majumdar for lively discussions of results and encouraging interest in the subject, to Peter Forrester and Mark Mezard for bringing a few relevant references to our attention, and to Ofer Zeitouni for informing us on his forthcoming rigorous large-deviation analysis of the problem. YF was supported by EPSRC grant EP/J002763/1 ``Insights into Disordered Landscapes via Random Matrix Theory and Statistical Mechanics''. PLD was supported by ANR Grant No.~09-BLAN-0097-01/2.

\end{document}